\documentclass[useAMS,usenatbib]{mn2e}
\usepackage{graphicx}
\def\note #1]{{\bf #1]}}

\title[Benefits of multiple sites for asteroseismic detections]{Benefits of multiple sites for asteroseismic detections}
\author[Arentoft et al.]
{T. Arentoft$^{1}$\thanks{E-mail:toar@phys.au.dk}, B. Tingley$^{1}$, 
J. Christensen-Dalsgaard$^{1}$,
H. Kjeldsen$^{1}$,
T. R. White$^{1,2,3}$ \newauthor
and F. Grundahl$^{1}$\\
$^{1}$Stellar Astrophysics Centre, Dept. of Physics and Astronomy, Aarhus University, Ny Munkegade 120, DK-8000 Aarhus C, Denmark\\
$^{2}$Sydney Institute of Astronomy (SIfA), School of Physics, University of Sydney, NSW 2006, Australia\\
$^{3}$Australian Astronomical Observatory, PO Box 915, North Ryde, NSW, 1670, Australia
}
\begin{document}

\date{Accepted. Received; in original form}

\pagerange{\pageref{firstpage}--\pageref{lastpage}} \pubyear{2002}

\maketitle

\label{firstpage}

\begin{abstract}
While {\it Kepler} has pushed the science of asteroseismology to limits
unimaginable a decade ago, the need for asteroseismic studies of
individual objects remains. This is primarily due to the limitations of 
single-colour intensity variations, which are much less sensitive to
certain asteroseismic signals. The best way to obtain the
necessary data is via very high resolution ground-based
spectrography. Such observations measure the perceived radial-velocity shifts that
arise due to stellar oscillations, which exhibit a much better
signal-to-noise ratio than those for intensity observations. 
SONG, a proposed network of 1\,m
telescopes with spectrographs that can reach R=110,000, was designed
with this need in mind. With one node under commissioning in
Tenerife and another under construction in China, an analysis of the 
scientific benefits of constructing additional nodes for the network is 
warranted. By convolving models of asteroseismic observables (mean
densities, small separations) with the anticipated window functions
for different node configurations, we explore the impact of the
number of nodes in the SONG network on the anticipated results, across
the areas of the HR diagram where solar-like oscillations are
found. We find that although time series from two SONG nodes, or in some 
cases even one node, will allow us to detect oscillations, 
the full 
SONG network, providing full temporal coverage, is needed for obtaining the 
science goals of SONG, including analysis of modes of spherical harmonic 
degree $l=3$.
\end{abstract}


\section{Introduction}
The {\it Kepler} mission has started a new and much more detailed chapter on
asteroseismology. Before launch, only a few stars exhibiting solar-like
oscillations had been studied with 
sufficient detail to allow the necessary analysis; however {\it Kepler} 
has made it possible to perform asteroseismology of thousands of stars
\citep{g1, chap1}, enabling the detection of not only acoustic modes but also 
of mixed modes in evolved stars \citep{b1}. 

While {\it Kepler} is a marvelous instrument for many aspects of
asteroseismology, it is limited in other respects. Stars exhibit more
intrinsic noise relative to their asteroseismic amplitudes in
intensities than in radial velocities. As such, it is possible to
detect oscillation frequencies to much higher precision and at
lower frequencies with radial-velocity
observations than using intensities. The lower-frequency modes have 
longer lifetimes, which in turn allow the frequencies to be measured with a 
higher degree of accuracy. Moreover, they allow us to 
study the helium abundance in the envelope via the ``acoustic glitch'', an
observable phenomenon which arises from helium ionization 
\citep{g3, h1}, and they are less affected by the near-surface effects in the 
stellar models \citep{cdt97}.

Furthermore, it is possible to detect $l=3$ modes through 
radial-velocity measurements, which supply further information about the 
stellar core \citep[e.g.,][]{cunmet07, jcd1}. This represents a major advantage of radial 
velocity time-series observations of solar-like oscillations, as intensity observations
can only make marginal detections of such modes, even from space.
We also note that radial-velocity measurements in general 
focus on much brighter stars than those observed by
{\it Kepler}, which means that more information about the stars is available
from other sources, such as interferometry, etc. 

Unfortunately, the analysis of radial-velocity modulations caused by solar-like 
oscillations requires ultra-high precision measurements, on the order of 
1 m\,s$^{-1}$ 
with time series of several days or weeks if not more in length.
Temporal coverage also has a strong impact on detection:
the lower the temporal coverage, the longer the time baseline
needs to be and the more pronounced the window function becomes.
Currently, only a very few
instruments are capable of reaching the necessary precision and these
are in high demand, making it difficult to obtain enough telescope
time to perform these studies.

SONG \citep[Stellar Observations Network Group,][]{song1, song2}
is an instrument designed for this very purpose. It is a proposed
network of robotic 1\,m telescopes with spectrographs capable of
reaching a radial-velocity precision of 1\,m\,s$^{-1}$ on stars down to 
magnitude $V=6$. Currently, SONG has one
node under commissioning in Tenerife and another under construction in
China, with additional nodes expected to be added in the future. It is
therefore very important for the project that we evaluate the impact of the
number of nodes on the scientific return of SONG. In this paper, we
explore how the number of nodes in SONG will affect the detection
of solar-like oscillations across regions 
of the HR diagram where such variations are expected,
focussing on stars on or near the main sequence.

\begin{figure}
 \includegraphics[width=84mm]{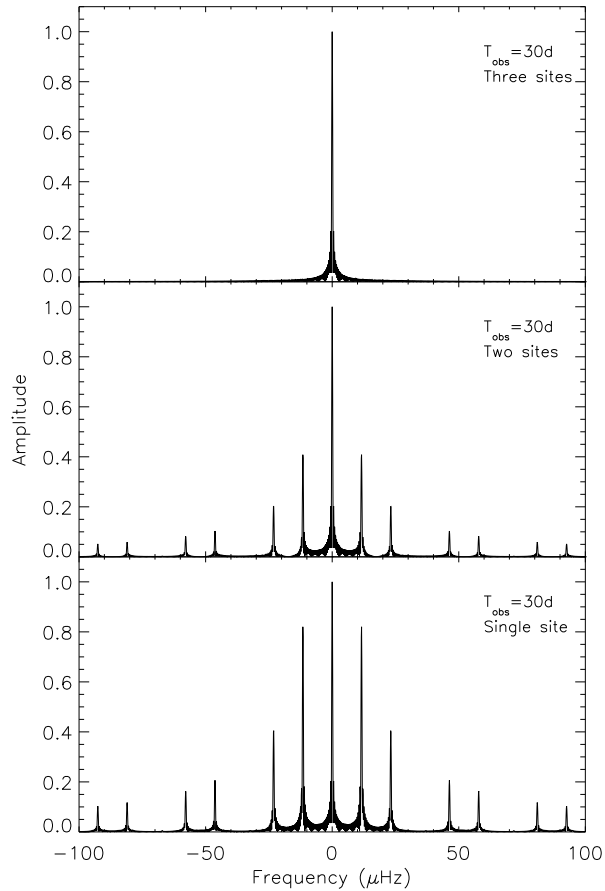}
\caption{Spectral window functions for 30 days of observations for
 24 (top), 16 (middle) and 8 (bottom) hours of time series coverage 
per 24 hours. The window functions represent time series
from three, two and one SONG nodes, respectively. 
The gaps in the window functions for the single- and two-site cases at
multiples of $3\,{\rm d}^{-1}$
are effects of the assumed length of night of 8h.}
\label{fig.windows}
\end{figure}

\begin{figure}
\hspace{-0.8cm} \includegraphics[width=95mm]{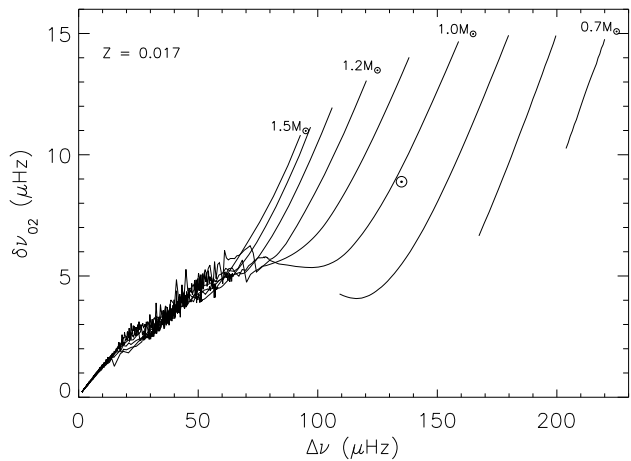}
\caption{The C-D diagram, showing the relationship between the large 
frequency separation $\Delta\nu$ and the small frequency separation 
$\delta\nu_{02}$ for stellar models with masses between $0.7-1.5M_\odot$ and
$Z=0.017$. The models are from \citet{w1}, see their paper for details.
The stars evolve from the top of the tracks and down toward lower values for
both $\Delta\nu$ and $\delta\nu_{02}$. The location of the Sun in this diagram
has been indicated by the solar symbol (note that the models have not been 
calibrated to match the Sun precisely).}
\label{fig.cdclean}
\end{figure}

\section[]{Method}

We study three different idealized configurations of nodes, each node
assumed to be observing for 8 hours: a single node (33\% temporal 
coverage), two nodes (67\% coverage) and three nodes (100\% coverage)
in order to investigate to what level the resulting spectral window functions 
which are shown in Fig. \ref{fig.windows}, cause 
overlap between different oscillation modes in the amplitude spectrum. 
Such overlaps are detrimental to the analysis of the
stellar oscillations, which is a
primary motivation for constructing a telescope network.

We use the asymptotic relation to predict the oscillation frequencies for 
different combinations of the large and small frequency separations across
the so-called C-D diagram, which plots the large and the small frequency 
separation against each other \citep{jcd84, jcd88, ulrich86}. 
We then use stellar models to combine the results from the analysis of 
overlapping peaks across the C-D diagram with two other, important 
observables; the mode lifetimes (which affect the mode linewidths)
and amplitudes of the solar-like oscillations. By doing so, we identify 
the areas in the HR diagram where solar-like oscillations can 
be observed and analysed with the different SONG-node configurations.
For clarity, we break this analysis into two parts: the first exploring the
impact of the window function on detection given a fixed mode linewidth, then
adding in the effects of realistic linewidths and amplitudes for stars of 
different ages and masses for the second part.

We focus on general trends in the present study, and disregard effects of 
weather, instrumental problems etc., which modify the window function and 
increase the noise level of the observations, as well as effects of rotation, 
which affects the linewidths of the frequency peaks. We also do not include
mixed modes, which is a complicating factor in the analysis of evolved stars. 

\subsection[]{The oscillation frequencies}

Before the impact of the window function on the detectability of solar-like
oscillations can be assessed, the basics of solar-like oscillations need
to be addressed.
Solar-like oscillations are acoustic oscillations which approximately 
follow the asymptotic relation \citep{v1, t1, g2}:
$$\nu_{n,l}\approx\Delta\nu(n+\frac{1}{2}l+\epsilon)-l(l+1)D_0,$$
where $n$ is the radial order, $l$ is the angular degree,
$\Delta\nu$ is the large frequency separation between modes of same $l$ 
(which measures the mean density of the star), $\epsilon$ 
is sensitive to the surface layers, and is set to $1$ in this work, and
$D_0 = \frac{1}{6}\delta_{02}$, where $\delta_{02}$ is the small frequency
separation between $l=0$ and $l=2$ modes (which is sensitive to the age 
of the star).
We assume in this study that all stars follow this 
relation, and we assume that the large and small frequency 
separations are constant across the entire frequency spectrum.

The frequency separations mentioned above are not enough by themselves to
assess the impact of the window function -- the analysis requires the
oscillation frequencies themselves. We calculate these using the asymptotic
relation for the values of $\Delta\nu$ and $\delta\nu_{02}$ covering the stellar
evolutionary tracks depicted in Fig.~\ref{fig.cdclean}, for modes with
$l=0-3$.

\begin{figure*}
\hspace{1.1cm} \includegraphics[width=130mm]{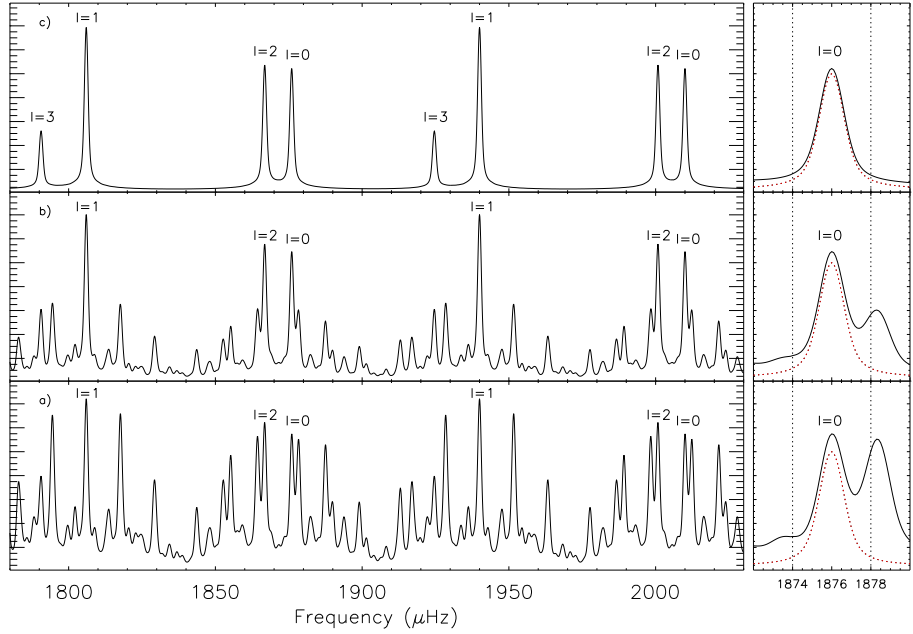}
\caption{Synthetic amplitude spectra with $\Delta\nu=134.0\, \mu$Hz and 
$\delta\nu_{02}=9.2 \,\mu$Hz (see text), observed from a single (a), 
from two (b) or from three (c) SONG sites. 
The rightmost panels display detailed views of a single mode, with
dashed red lines representing the undisturbed oscillation mode,
black lines representing the modelled amplitude spectrum,
which includes contributions from multiple modes as described by
the asymptotic relation, and vertical dotted lines showing
the spectral region used to calculate the level of contamination.
The contamination of the synthetic amplitude spectrum in the bottom panel
is below 30\%, while it is below 10\% in both the center and top panels.
The positions of the $l=3$ modes are indicated in the top plot only,
for clarity.
}
\label{fig.ShowOver}
\end{figure*}

\begin{figure*}
 \includegraphics[width=130mm]{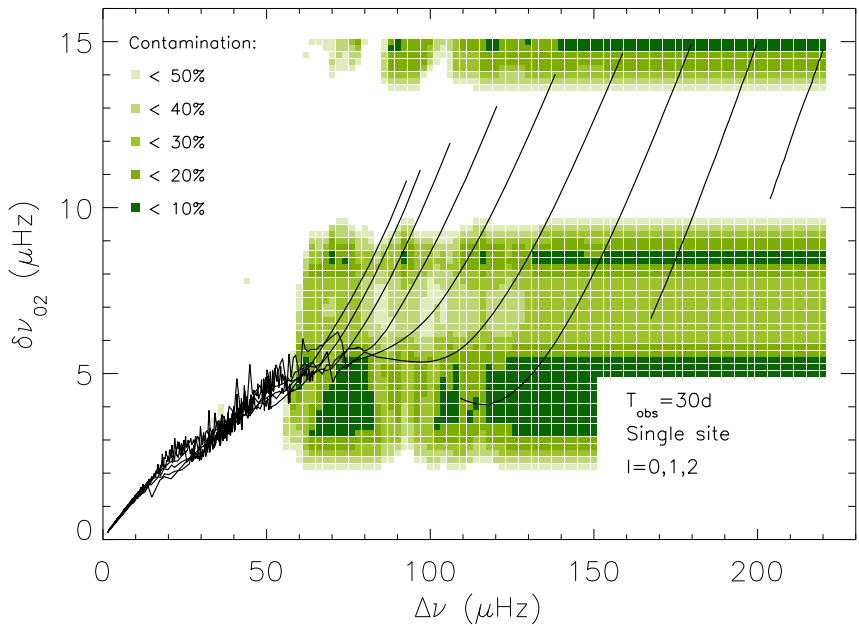}
\caption{
The C-D diagram colour-coded to illustrate amount of overlap between 
peaks based on each combination of ($\Delta\nu$, $\delta\nu_{02}$), the 
asymptotic relation and the spectral window function corresponding to 
observations from a single site. 
Modes of $l=0,1,2$ have been included when calculating the contamination
levels.
Dark green areas correspond to combinations of ($\Delta\nu$, $\delta\nu_{02}$)
which result in very little overlap, 
while light-green areas correspond to combinations with significant overlap. 
White areas correspond to overlaps higher than 50\% in amplitude, except in 
the upper left and lower right corners where no calculations have been done.
The stellar models are, as in Fig.~\ref{fig.cdclean}, from \citet{w1},
again with masses between $0.7-1.5M_\odot$ and $Z=0.017$. The width of the 
frequency peaks has been set to 1.5\,$\mu$Hz.
}
\label{fig.surfaceSS}
\end{figure*}

\subsection[]{Constructing the amplitude spectra}

To facilitate the discussion of the mode overlap we introduce modified
window functions $w_{{\rm mod}, k}(\nu)$, where $k = 1, 2, 3$ correspond
to the number of nodes.
These are defined by modifying the original window functions to 
reflect peaks whose intrinsic width $2/T \approx 0.77\, \mu$Hz is determined
by the assumed length of 30 days of the time series, combined
with a Lorentzian with a width determined by the mode lifetime.
Here we have assumed a mode lifetime for solar-like oscillations
similar to that of the Sun (3 days) -- a reasonable simplification
for the purpose of studying the impact of the spectral
window functions.
The relationship between the mode lifetime
($\tau$) and the natural linewidth $\Gamma$ of the mode yields
$\Gamma = 1/(\pi \tau) \approx 1.23\, \mu$Hz.
In practice we constructed $w_{{\rm mod},k}$ by an iterative process, where
we smooth the original window function until it has a full width at half 
maximum (FWHM) of $1.5\, \mu$Hz. This corresponds roughly to adding the 
intrinsic width of the window function
and the natural mode linewidth in quadrature. The peak amplitudes of the 
modified window functions are kept at 1.0.

When constructing the amplitude spectra, we also use the fact that modes 
with different $l$-values have different spatial response functions, or
sensitivities, for line-of-sight Doppler observations. These are denoted $S_l$, 
and lead to perceived differences in amplitude \citep[e.g.,][]{b0};
the sensitivity of $l=0$ is set to 1.00, $l=1$ to 1.36, $l=2$ to 1.03 and
$l=3$ to 0.48.\\ 

We now construct the amplitude spectra $a_k(\nu)$ by combining
the frequencies, $\nu_{n,l}$, with the modified window functions,
and taking into account the $l$-dependent sensitivities, $S_l$:
\begin{equation}
a_k(\nu) = \sum_{n,l} S_l \, w_{\rm mod,k}(\nu - \nu_{n,l}) \; .
\end{equation}
Note that by simply adding the amplitude contributions from each mode we 
ignore the interference between the modes, depending on their relative 
phases, which would affect a more realistic combination of the contributions 
in Fourier space (or the time domain).
We expect that this has little effect on the present simplified analysis.

In Fig.~\ref{fig.ShowOver}, we show 
amplitude spectra for $\Delta\nu = 134.0\,\mu$Hz and $\delta\nu_{02} = 9.2\,\mu$Hz,
as observed from one, two or three SONG telescopes. 
From these plots, we can see that the artificial peaks introduced by 
the spectral window function of neighbouring modes can cause the different 
modes to interfere with each other, as one would expect.

\subsection[]{The contamination of the modes}

The degree of overlap (or {\it contamination}) is an important parameter in
this study. In this part, we examine the amount of contamination arising 
due to the spectral window function, using the fixed FWHM of 1.5\,$\mu$Hz
for the mode peaks.
In order to evaluate the contamination for a given mode at frequency
$\nu$, we determine the signal in the amplitude spectrum, within a 4\,$\mu$Hz 
wide frequency band (see Fig.~\ref{fig.ShowOver}), and compare it to the 
signal which originates from the undisturbed mode.

From the assumed signal in equation (1) we can,
for the different $l$-values of the modes, determine the amount of signal 
$a_{l,k}$ (including contamination from other modes) in a mode with frequency 
$\nu_l$ as
\begin{equation}
a_{l,k} = \int_{\nu_1}^{\nu_2} \! a_k(\nu)\,\mathrm{d\nu} \; .
\end{equation}
Here, $\nu_1 = \nu_l - 2 \,\mu \rm{Hz}$ and $\nu_2 = \nu_l + 2\,\mu \rm{Hz}$. 
Examples of amplitude spectra $a_k(\nu)$ were shown for 1, 2 and 3 sites in 
Fig.~\ref{fig.ShowOver}.
The spectra repeat themselves for each radial order $n$, so for the $l=0$ 
modes, we may use the mode at $1876 \,\mu \rm{Hz}$, for determining the total 
amount of signal. In this case, $\nu_1 = 1874\,\mu \rm{Hz}$ and 
$\nu_2 = 1878\,\mu \rm{Hz}$. 

In the same way, we can obtain the amount of signal originating only from 
the mode in question, i.e., the signal that would be inside the frequency 
band, if the mode were undisturbed.
From the central peak in the modified window function $w_{\rm mod,k}(\nu)$
discussed above the amount of signal $w_{l,k}$ for an undisturbed mode
is found as 
\begin{equation}
w_{l,k} = \int_{-2\,\mu \rm{Hz}}^{+2\,\mu \rm{Hz}} \! S_l\, w_{\rm mod,k}(\nu)\,\mathrm{d\nu} \;.
\end{equation}

In this way, we obtain for each node configuration, and for each $l$-value, 
the amount of signal in the combined amplitude spectrum, $a_{l,k}$,
and the amount of signal in the undisturbed mode, $w_{l,k}$.
We use these values to calculate the amount of contamination as
\begin{equation}
c_{l,k} = (a_{l,k} - w_{l,k}) / w_{l,k} \; .
\end{equation}
For each node configuration, and for each set of 
($\Delta\nu, \delta\nu_{02}$)-values, we first determine the 
contamination for
$l = 0, 1, 2$, leaving out $l=3$, and we assign the highest 
of these three numbers as the 
contamination level for that set of parameters. For example, if for one 
of the node configurations, and for a given combination of $\Delta\nu$ and 
$\delta\nu_{02}$, the $l=0$ modes are contaminated by 5\%, $l=1$ 
by 15\% and $l=2$ by 25\%, we assign a contamination level $c_k$ for that 
specific combination of parameters of 25\%. 

When calculating the contamination, we exclude the contribution
from the wings of distant modes; these cause an offset 
in the spectra, manifesting as a slightly increased background level
but having little effect on asteroseismic analysis. This effect is exemplified
by the mode depicted in the upper-right
panel of Fig.~\ref{fig.ShowOver}; despite some small amount of blending,
it is effectively isolated, with a contamination of 0\%.

\begin{figure}
\hspace{-0.8cm} \includegraphics[width=95mm]{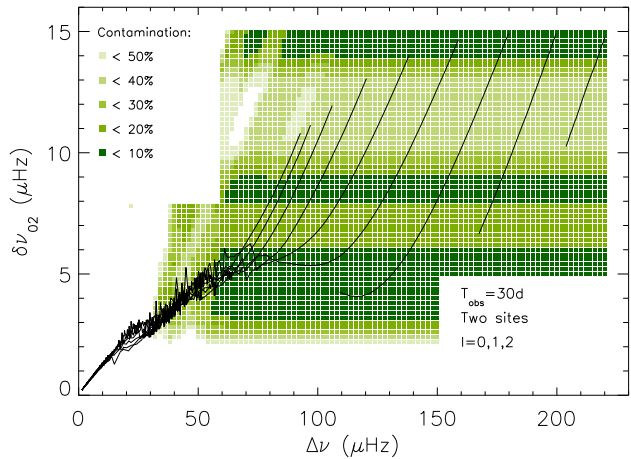}
\caption{
Same as Fig.~\ref{fig.surfaceSS}, but for two SONG telescopes.
}
\label{fig.surfaceDS}
\end{figure}

\begin{figure}
\hspace{-0.8cm} \includegraphics[width=95mm]{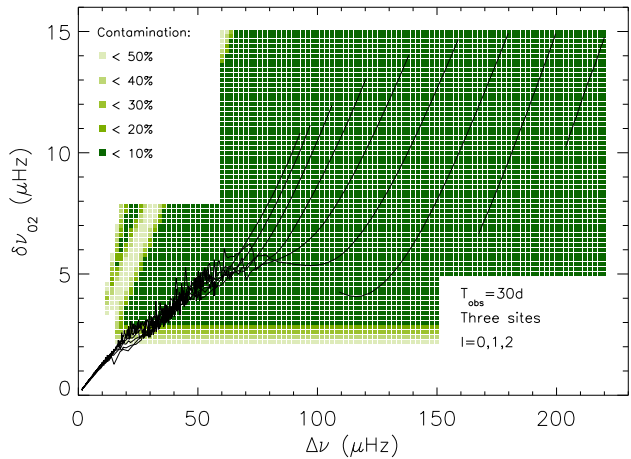}
\caption{
Same as Fig.~\ref{fig.surfaceSS}, but for three SONG telescopes. 
}
\label{fig.surfaceTS}
\end{figure}

In order to visualize the effect of contamination due to the window function, 
we compare the $c_k$-values to 5 different levels: below 10\%, 20\%, 30\%, 
40\% and 50\% contamination in amplitude, with lower
values corresponding to lower degrees of overlap between modes.
In the analysis, we cover the parameter space in the C-D diagram with a grid 
with steps of 2\,$\mu$Hz in $\Delta\nu$ and 0.2\,$\mu$Hz in $\delta\nu_{02}$. 
The results of this analysis for 1, 2 and 3 nodes in the network, 
excluding the lower-amplitude $l=3$ modes from the computations, 
can be seen in Figs.~\ref{fig.surfaceSS}--\ref{fig.surfaceTS}. 

\begin{figure}
\hspace{-0.8cm} \includegraphics[width=95mm]{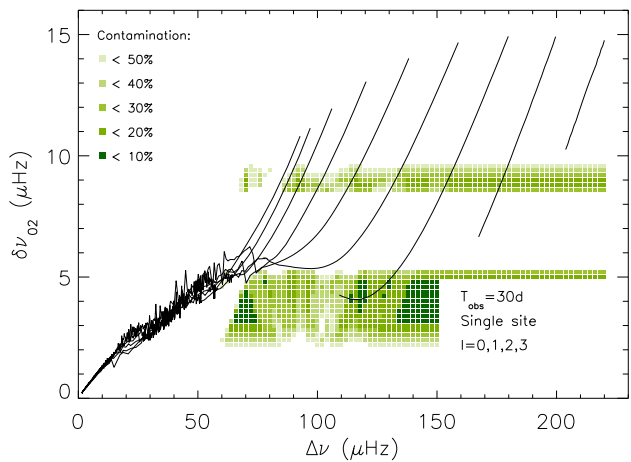}
\caption{Same as Fig.~\ref{fig.surfaceSS} (a single SONG node), 
with $l=3$ modes included in the calculation of the mode contamination.
Including these modes results in models with $\delta\nu_{02}\sim 7 \,\mu$Hz
becoming excluded. For these models, $\delta\nu_{13}\sim 11.5 \,\mu$Hz
(corresponding to 1 d$^{-1}$), causing overlap between $l=1$ and 
$l=3$ modes.
}
\label{fig.surfaceSS-l3}
\end{figure}

\begin{figure}
\hspace{-0.8cm} \includegraphics[width=95mm]{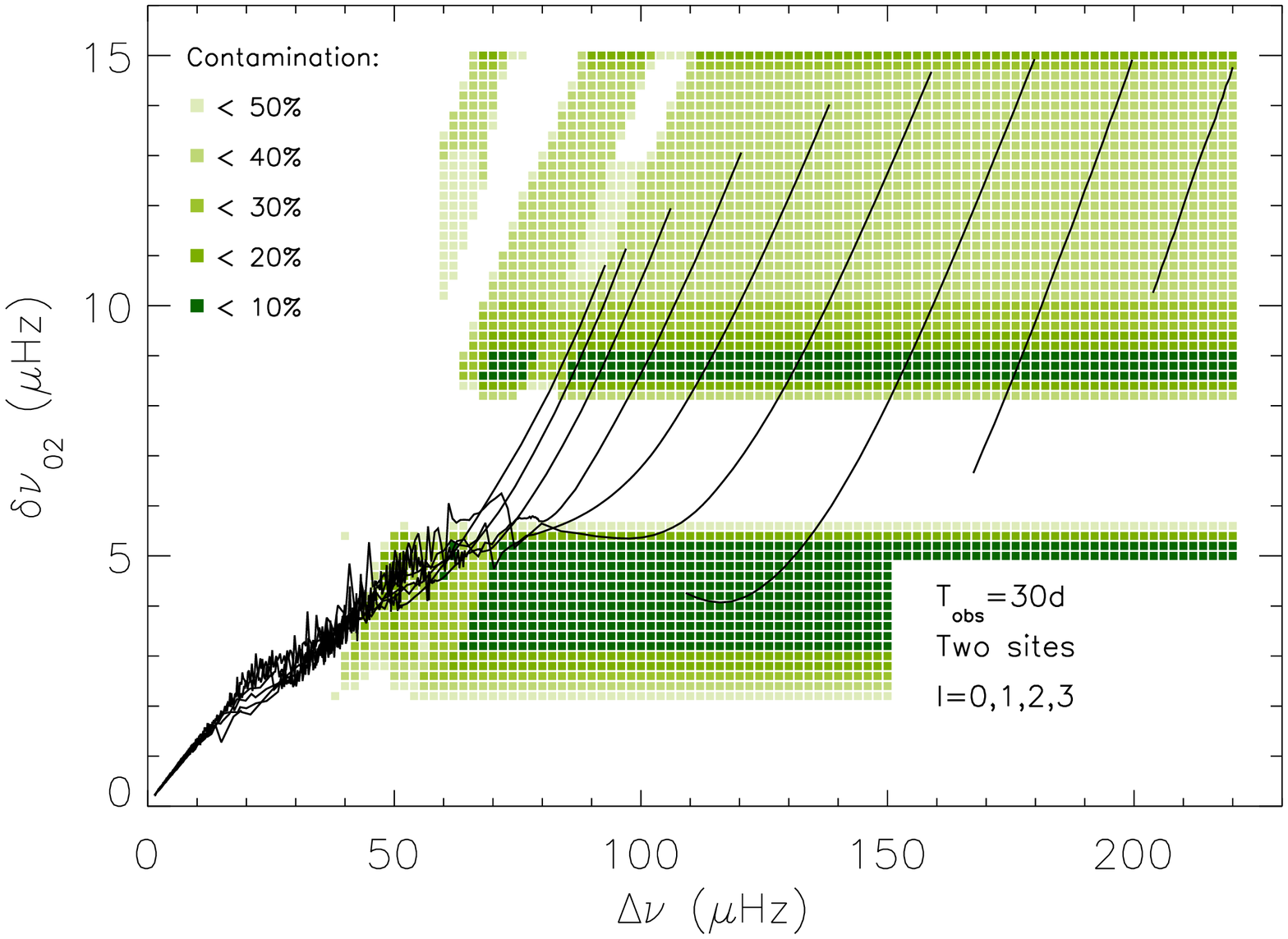}
\caption{Same as Fig.~\ref{fig.surfaceSS-l3}, but for two SONG telescopes.
}
\label{fig.surfaceDS-l3}
\end{figure}

We then extended the calculations of the mode contamination
to include the $l=3$ modes. The results are seen in 
Figs.~\ref{fig.surfaceSS-l3} and \ref{fig.surfaceDS-l3}.
The $l=3$ modes are separated from $l=1$ modes by 
$\delta\nu_{13} = \frac{5}{3}\delta\nu_{02}$ according to the asymptotic 
relation. This leads to the exclusion of more cases due to mode-peak overlap.
 
\begin{figure*}
\includegraphics[width=130mm]{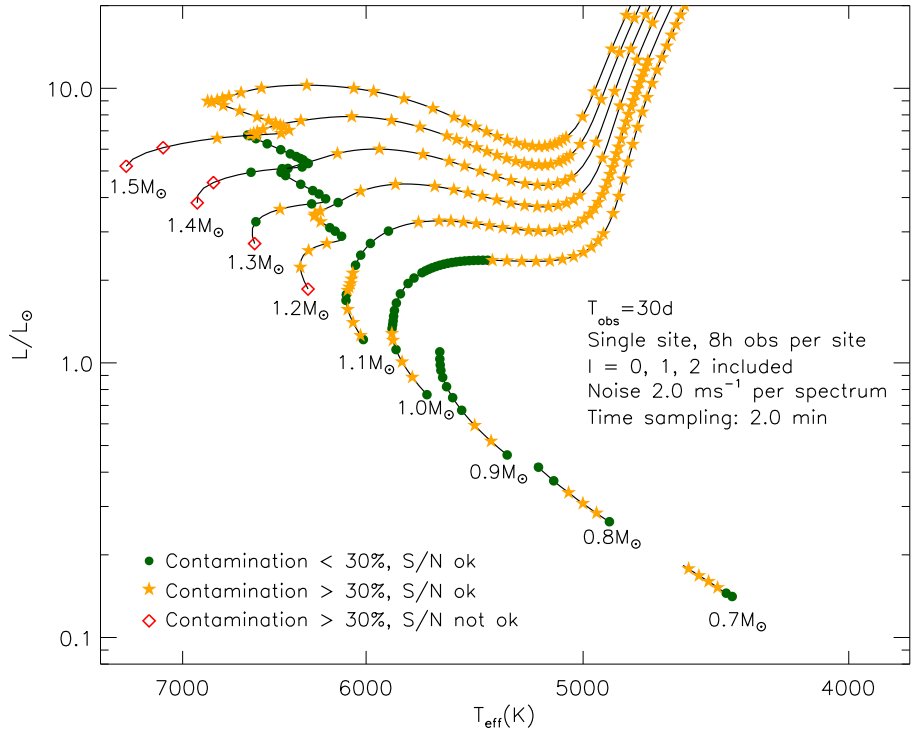}
\caption{Colour-coded HR diagram showing the results of our analysis for 
contamination in the amplitude spectrum of up to 30\% for a single
SONG node and a time series covering one month,
for modes with frequencies near the estimated frequency $\nu_{\rm max}$ of
maximum amplitude.
Individual models are marked according to the analysis, with green circles
marking the models for which seismic analysis is possible, based on our 
criteria, 
yellow stars marking models with contamination above 30\% due to
either the nature of the frequency structure as defined by the asymptotic relation
or large mode linewidths, and red diamonds marking models that fail on both
the amplitude and overlap requirements. Note that this and the 
following plots assume a certain noise level per spectrum, and hence do not 
contain any information on the apparent magnitude of real-sky targets. 
}
\label{fig.codedSS}
\end{figure*}

\subsection[]{Detecting the seismic signals}

We now combine the analysis above with the effects 
of oscillation mode amplitudes and linewidths on the 
detection of the asteroseismic signals. In short, the amplitudes must be 
sufficiently high compared with the noise level for SONG to allow 
secure detection of the oscillations and the mode linewidths
must be narrow enough to allow separation -- for some 
stars, the linewidths can be comparable to or even exceed
the small frequency separation, causing significant overlap between modes 
even for a 100\% duty cycle. 
We base the analysis on estimated parameters near the frequency $\nu_{\rm max}$
of maximum amplitude.

We use the theoretical stellar models to estimate both the oscillation
amplitudes and the mode linewidths (i.e., $\Gamma$) across 
the HR diagram. The mode
linewidths depend on effective temperature \citep{a2} and 
can therefore vary greatly for different types of stars -- indeed, 
asteroseismology can
be extremely difficult for stars with linewidths similar
to their frequency separations \citep[e.g.,][]{a3}.
The mode linewidths are estimated using the relation between $\Gamma$ 
and the effective temperature, at maximum mode height,
from equation (2) and Table 2 in \citet{a2}:

\begin{equation}
\Gamma = 0.46\,\mu{\rm Hz} + 0.75\,\mu{\rm Hz}\left(\frac{T_{\rm eff}}{5777\,{\rm K}}\right)^{15.4} \; .
\end{equation}
We extrapolate this relation to include stars with temperatures
above $7000\,$K, even though the calibration is based on stars
no hotter than $6400\,$K \citep{a2}. 
By doing so, we slightly overestimate the mode linewidths for these more 
massive stars relative to those found for F stars in \citet{w2}.
We assume that we cannot detect the seismic signal for stars with mode 
linewidths which are half the value of either the small or large frequency 
separation. For these stars, the degree of overlap of the mode
frequencies compromises the asteroseismic analysis. This is, however, a 
marginal effect in our analysis. 
 
The overall oscillation amplitudes are estimated for the stellar models 
by scaling 
from the solar value, based on equation (17) in \citet{kb1}:
\begin{equation}
A_{\rm vel}\approx 0.23\,\, {\rm m\, s^{-1}} \times \frac{(L/{\rm L}_\odot)(\tau_{\rm osc}/\tau_{\rm osc,\odot})^{0.5}}{(M/{\rm M}_\odot)^{1.5}(T_{\rm eff}/5777{\, \rm K})^{2.25}} \; ,
\end{equation}
where $\tau_{\rm osc}$ is the mode lifetime. 

For this analysis, we compare these amplitudes to the expected noise level
for SONG. This depends on the brightness of the star being observed,
its spectral type and class, its rotational velocity, and other secondary factors. 
Here we use a fixed noise value of 2\,m\,s$^{-1}$ per spectrum,
based on estimates of the properties of the SONG observations and
preliminary data from the SONG prototype, with a observing
cadence of one datapoint per two minutes, 8 hour long nights per site,
and a time series covering 30\,d. With this, we expect to achieve
noise levels in the amplitude spectra of 
4.2\,cm\,s$^{-1}$ for a single SONG-node, 3.0\,cm\,s$^{-1}$ for two nodes, 
and 2.4\,cm\,s$^{-1}$ for three nodes. We consider the seismic signal 
detectable if the 
estimated oscillation amplitude is at least 4 times higher than the 
predicted noise level. These S/N estimates are valid near the frequency of 
maximum amplitude for each model, $\nu_{\rm max}$.
Figs.~\ref{fig.codedSS}--\ref{fig.coded3DS} show the results of this analysis.

\section[]{Results}

Figs.~\ref{fig.surfaceSS}--\ref{fig.surfaceDS-l3} show the results for the 
analysis of the impact of window functions on detecting asteroseismic signals for
a fixed linewidth of 1.5\,$\mu$Hz. From these figures, we can see that
stars with $\delta\nu_{02}$ close to 11.57\,$\mu$Hz (which corresponds to
one cycle per day) will show
significant overlap between modes when observed from a single site.
The width of the horizontal band centered at 11.57\,$\mu$Hz in
$\delta\nu_{02}$ depends on the assumed mode linewidth of
1.5\,$\mu$Hz; a smaller or larger assumed linewidth means a smaller
or larger area being excluded for $\delta\nu_{02}$ values near 11.57\,$\mu$Hz.

From Fig.~\ref{fig.surfaceSS}, we can see that even for single-site observations, 
there are areas in the C-D diagram where the mode frequencies will not 
overlap, despite the significant side-lobes in the spectral window function. 
This is the case for stars slightly more evolved and less massive than the
Sun, among others. This figure also demonstrates the sensitivity of asteroseismology
from a single-site to stellar properties: some stars can be targeted from a
single site without the analysis being hampered by confusion from
overlapping peaks in the amplitude spectrum, while other stars with just
slightly different properties cannot.

Figs.~\ref{fig.surfaceDS} and \ref{fig.surfaceTS} show the results 
of the analysis for two and three SONG sites. For two sites
(Fig.~\ref{fig.surfaceDS}), the amplitude spectra of stars with small 
separations near 11.57\,$\mu$Hz will display a 40\% overlap between 
$l=0$ and $l=2$ modes, in good agreement with the corresponding spectral 
window function in Fig.~\ref{fig.windows}. For three sites, problems with 
overlapping peaks are eliminated, illustrating the benefits of full temporal coverage. 
When extending the analysis to include $l=3$ nodes
(Figs.~\ref{fig.surfaceSS-l3} and \ref{fig.surfaceDS-l3}),
we find that full temporal coverage is required for the detection of
$l=3$ modes for most types of stars.

\begin{figure}
\hspace{-0.8cm} \includegraphics[width=95mm]{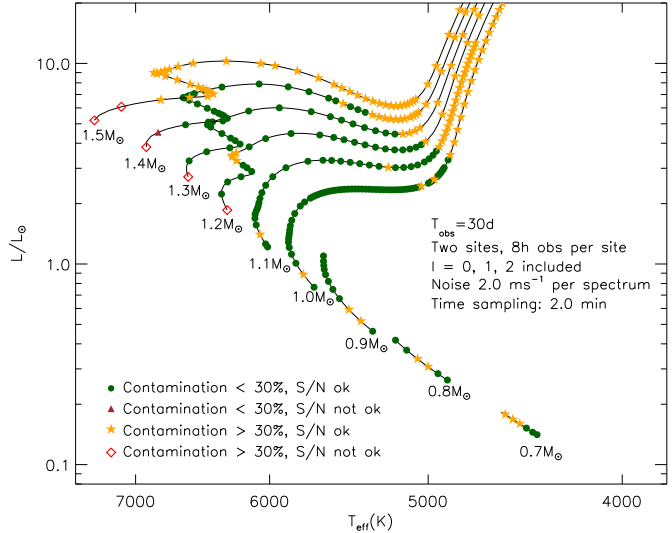}
\caption{
Same as Fig.~\ref{fig.codedSS}, but for observations with two SONG-nodes.
A brown triangle marks a model for which the the signal-to-noise (S/N) of the 
oscillations is below 4.  
\label{fig.codedDS}
}
\end{figure}

\begin{figure}
\hspace{-0.8cm} \includegraphics[width=95mm]{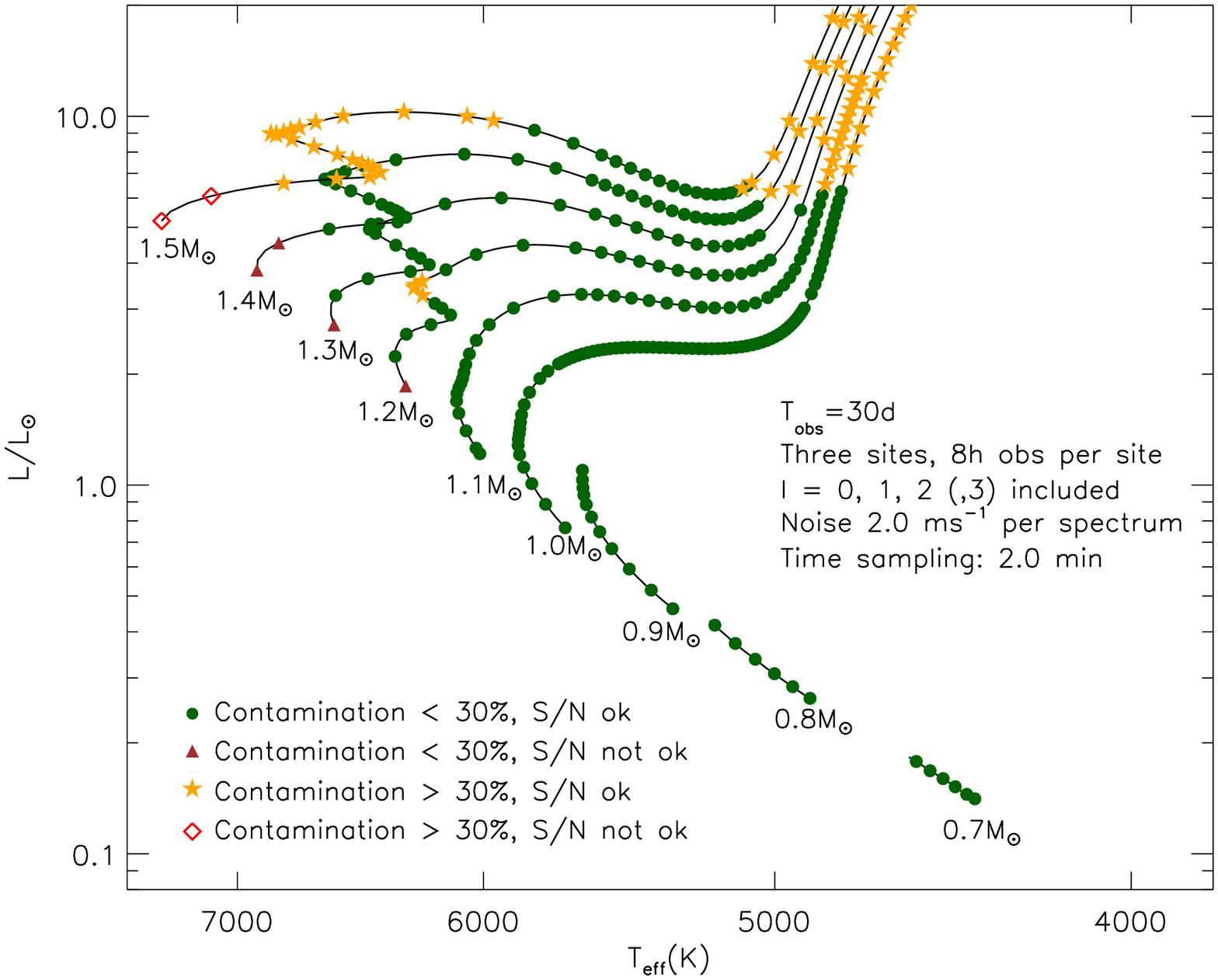}
\caption{
Same as Fig.~\ref{fig.codedSS}, but for observations with three SONG-nodes.
The figure does not change if $l=3$ modes are included in the analysis.
}
\label{fig.codedTS}
\end{figure}

\begin{figure}
\hspace{-0.8cm} \includegraphics[width=95mm]{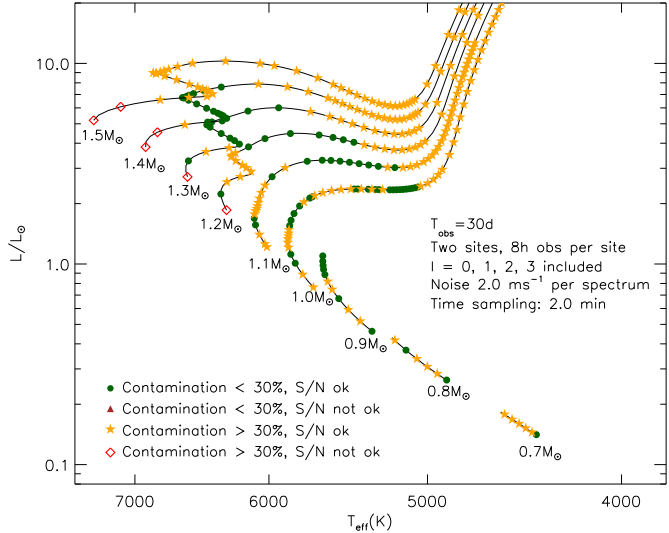}
\caption{
Same as Fig.~\ref{fig.codedDS} (observations with two SONG-nodes), 
but with $l=3$ modes included in the analysis.} 
\label{fig.coded3DS}
\end{figure}

\begin{figure}
\includegraphics[width=70mm]{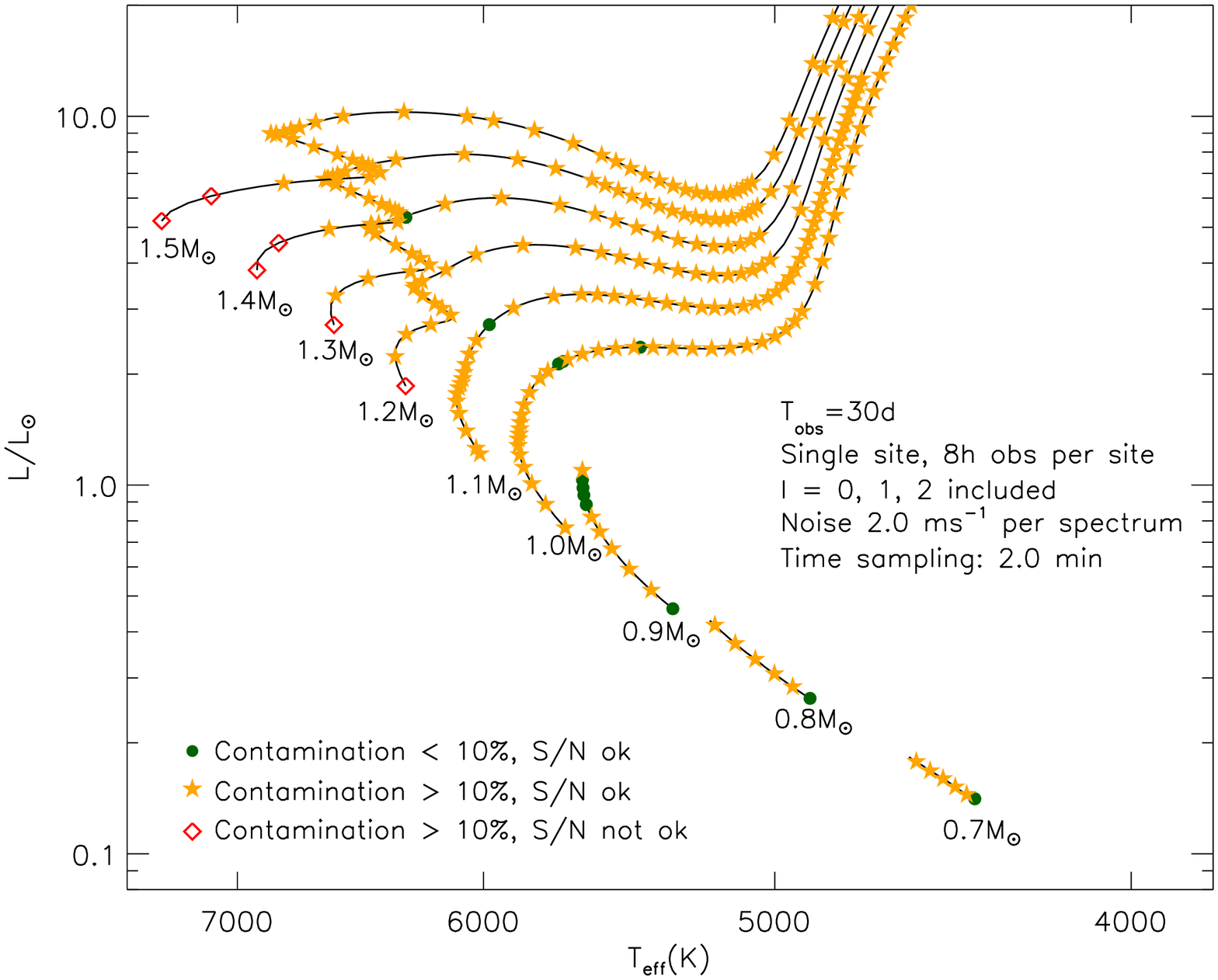} \\
\includegraphics[width=70mm]{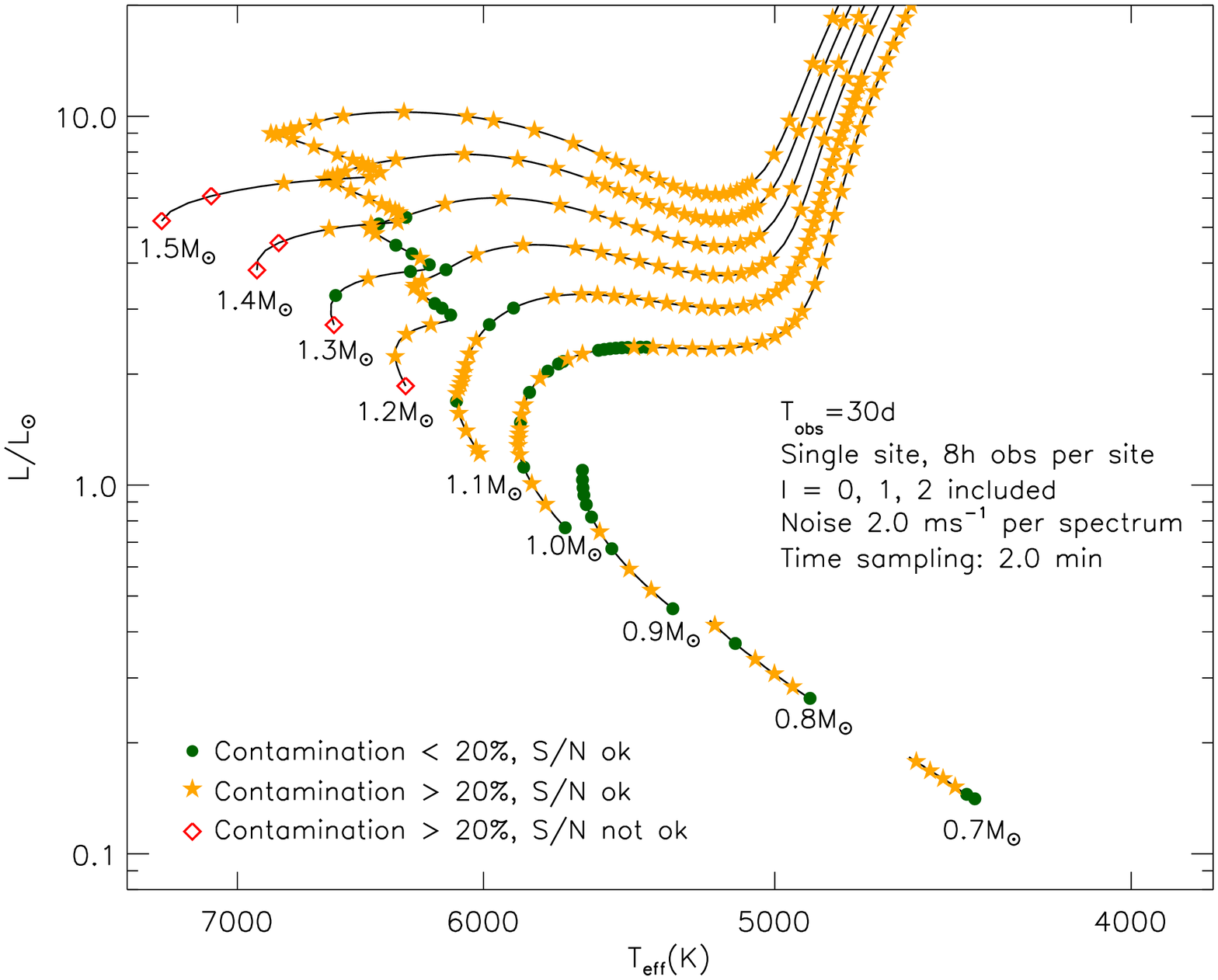} \\
\includegraphics[width=70mm]{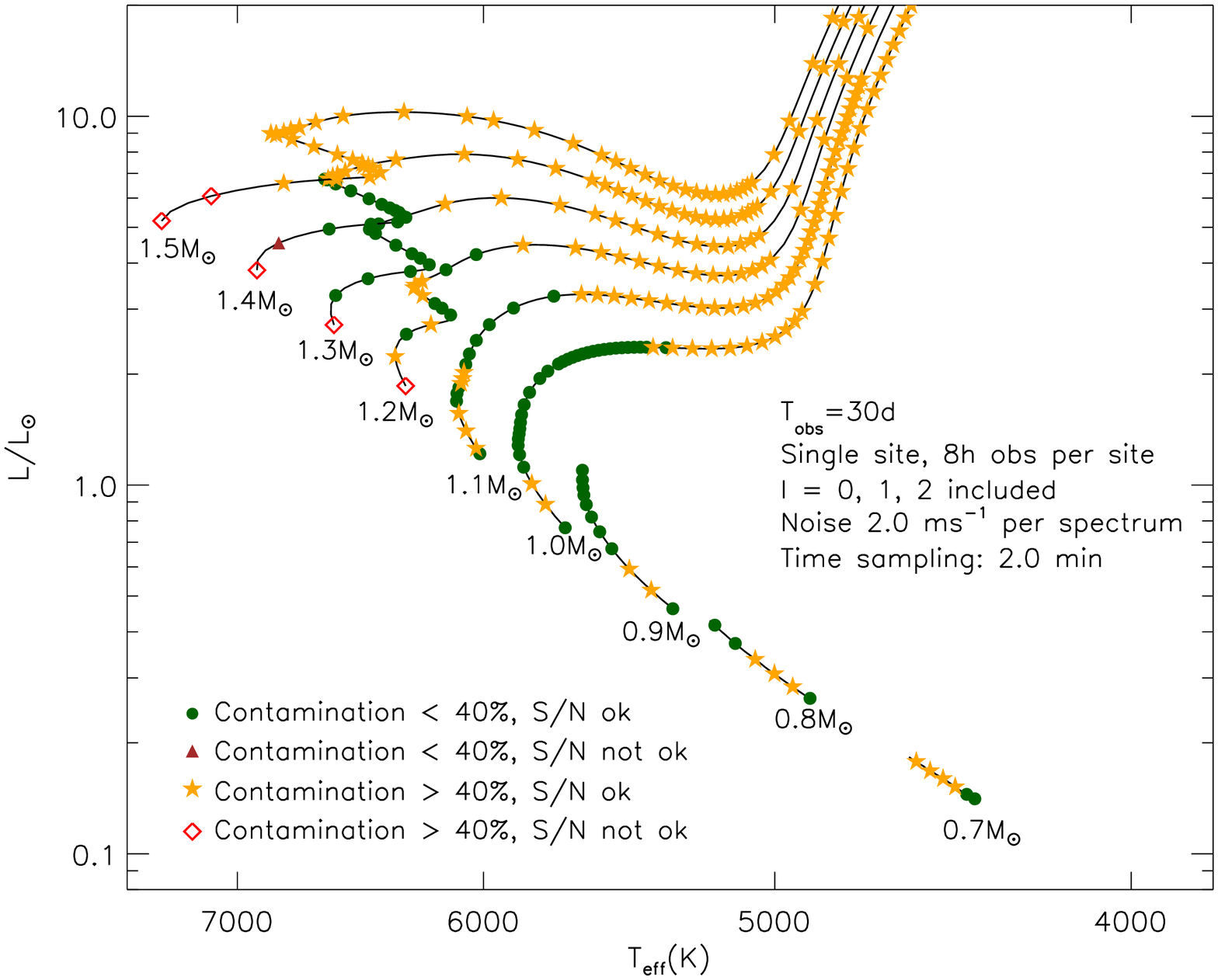} \\
\includegraphics[width=70mm]{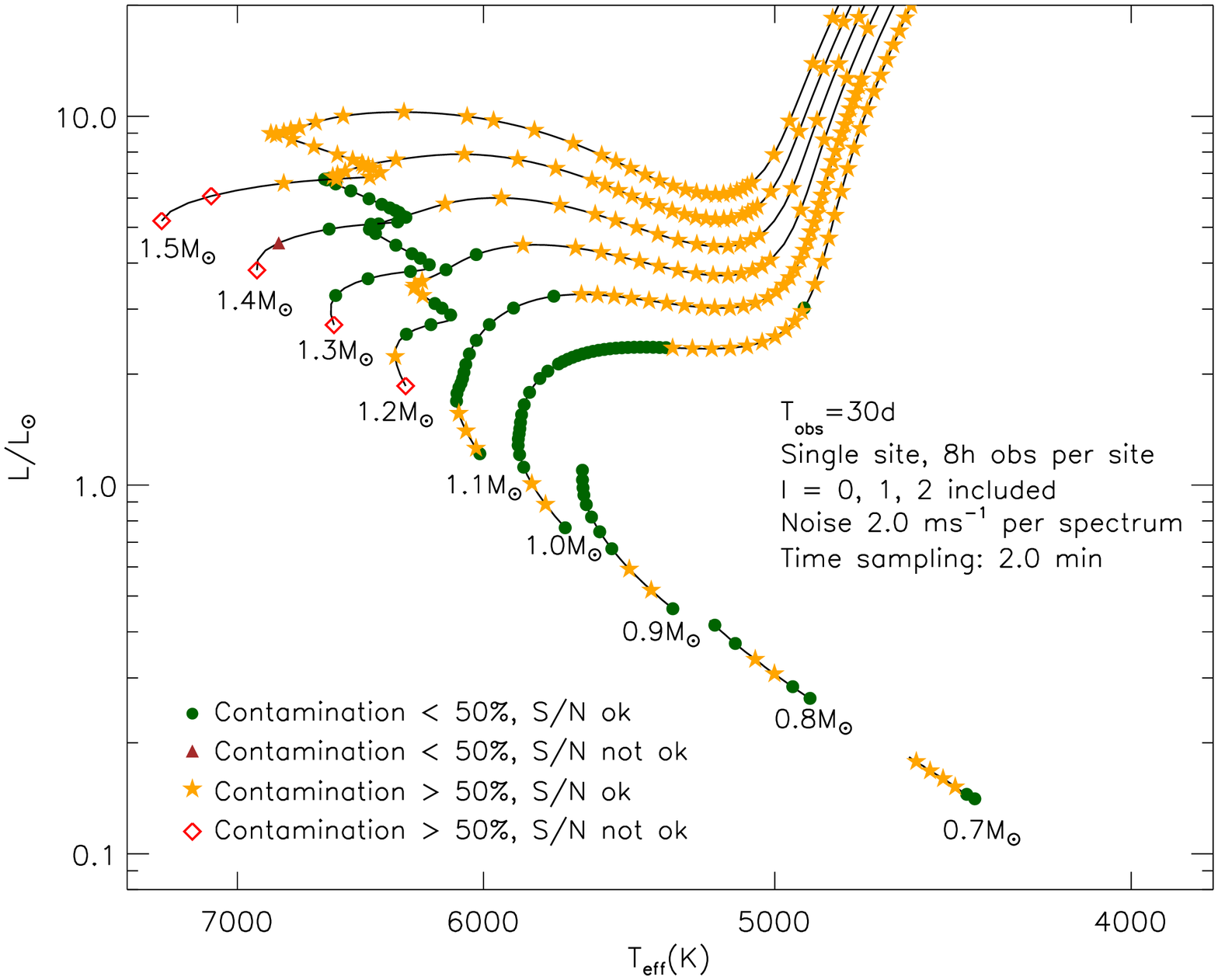} \\
\caption{Same as Fig.~\ref{fig.surfaceSS}, for a single site including 
$l=0-2$ and acceptable 
contamination levels of 10, 20, 40 and 50 per cent.} 
\label{fig.SS2}
\end{figure}

\begin{figure}
\includegraphics[width=70mm]{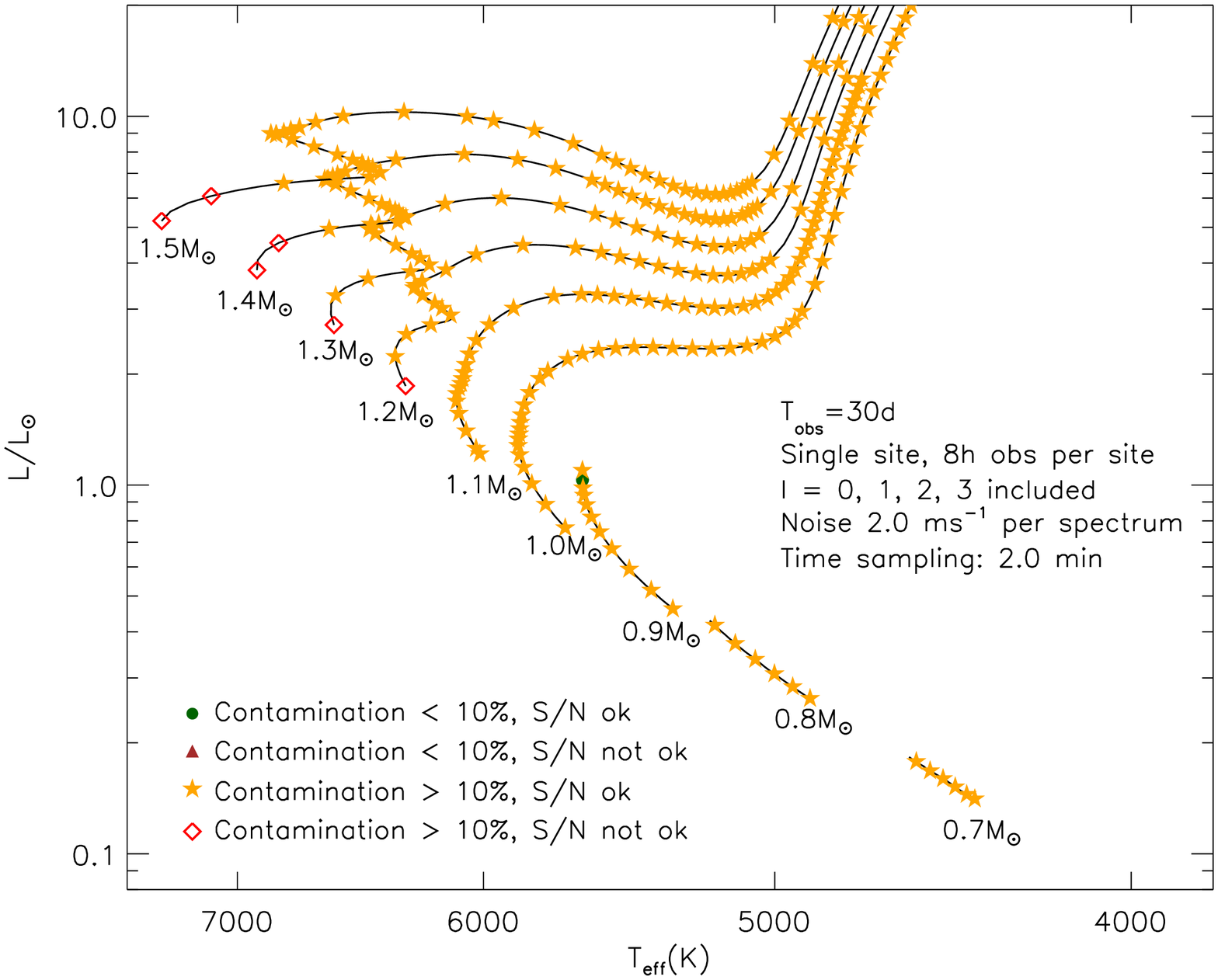} \\
\includegraphics[width=70mm]{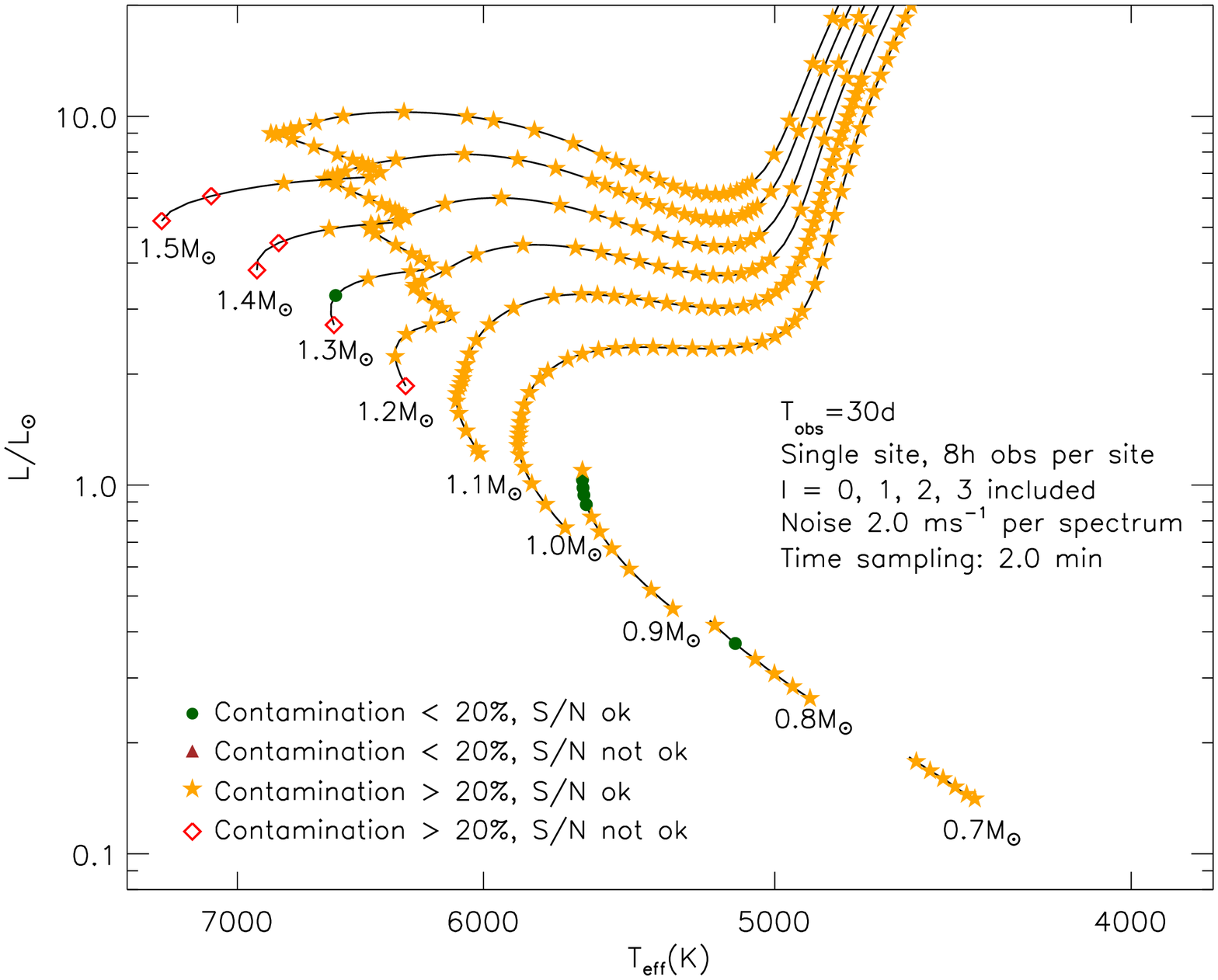} \\
\includegraphics[width=70mm]{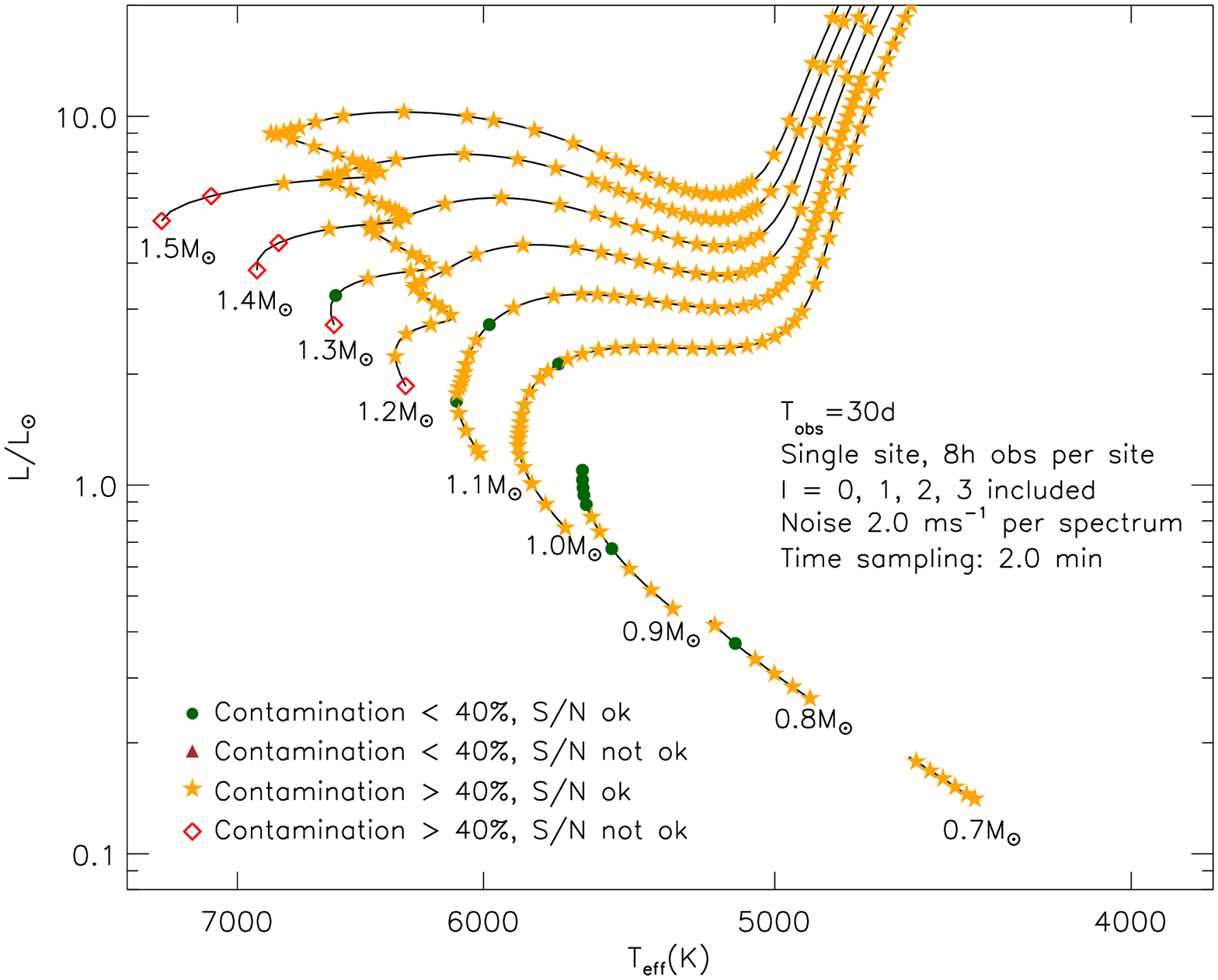} \\
\includegraphics[width=70mm]{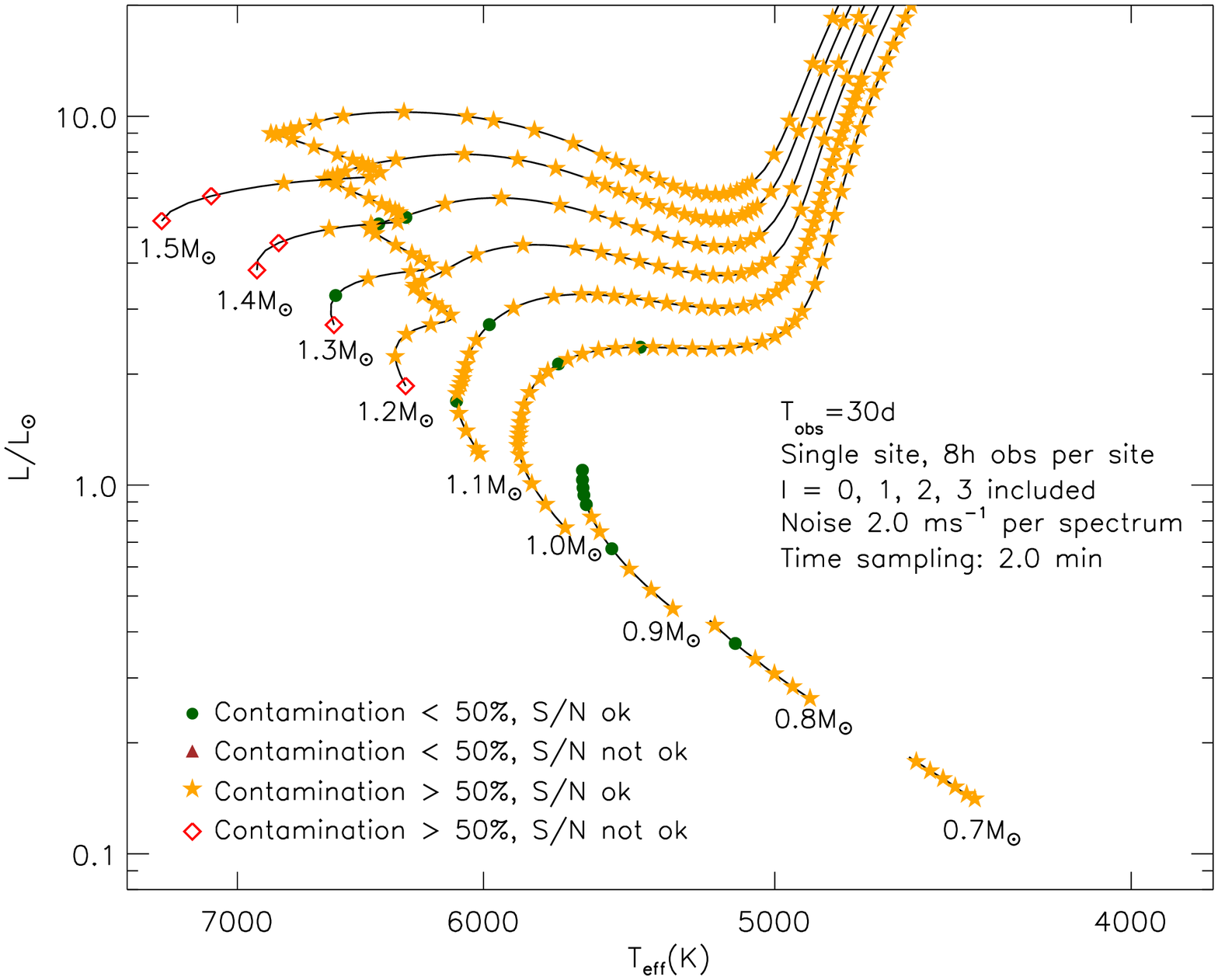} \\
\caption{Same as Fig.~\ref{fig.surfaceSS}, for a single site including 
$l=0-3$ and acceptable 
contamination levels of 10, 20, 40 and 50 per cent.} 
\label{fig.SS3}
\end{figure}

\begin{figure}
\includegraphics[width=70mm]{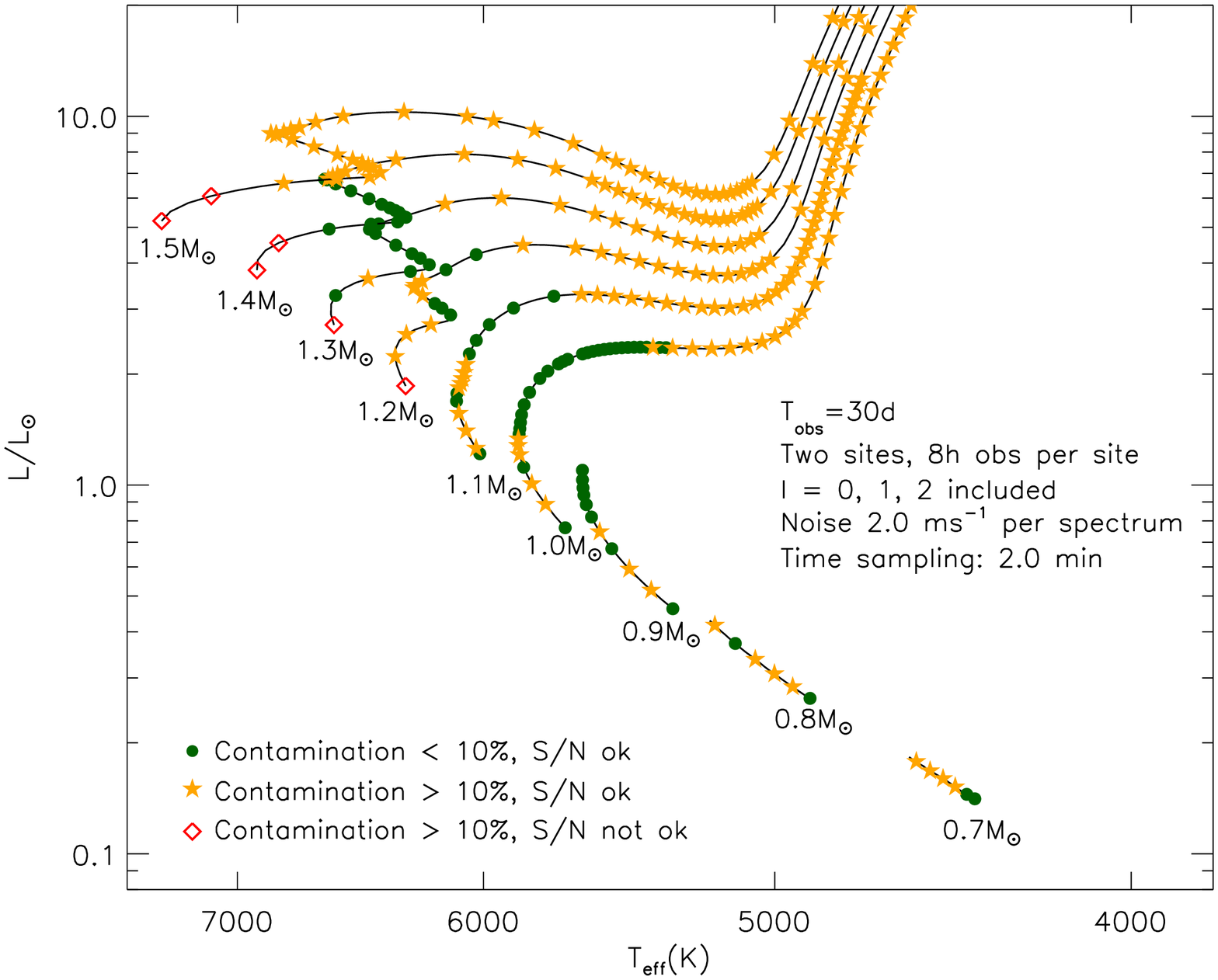} \\
\includegraphics[width=70mm]{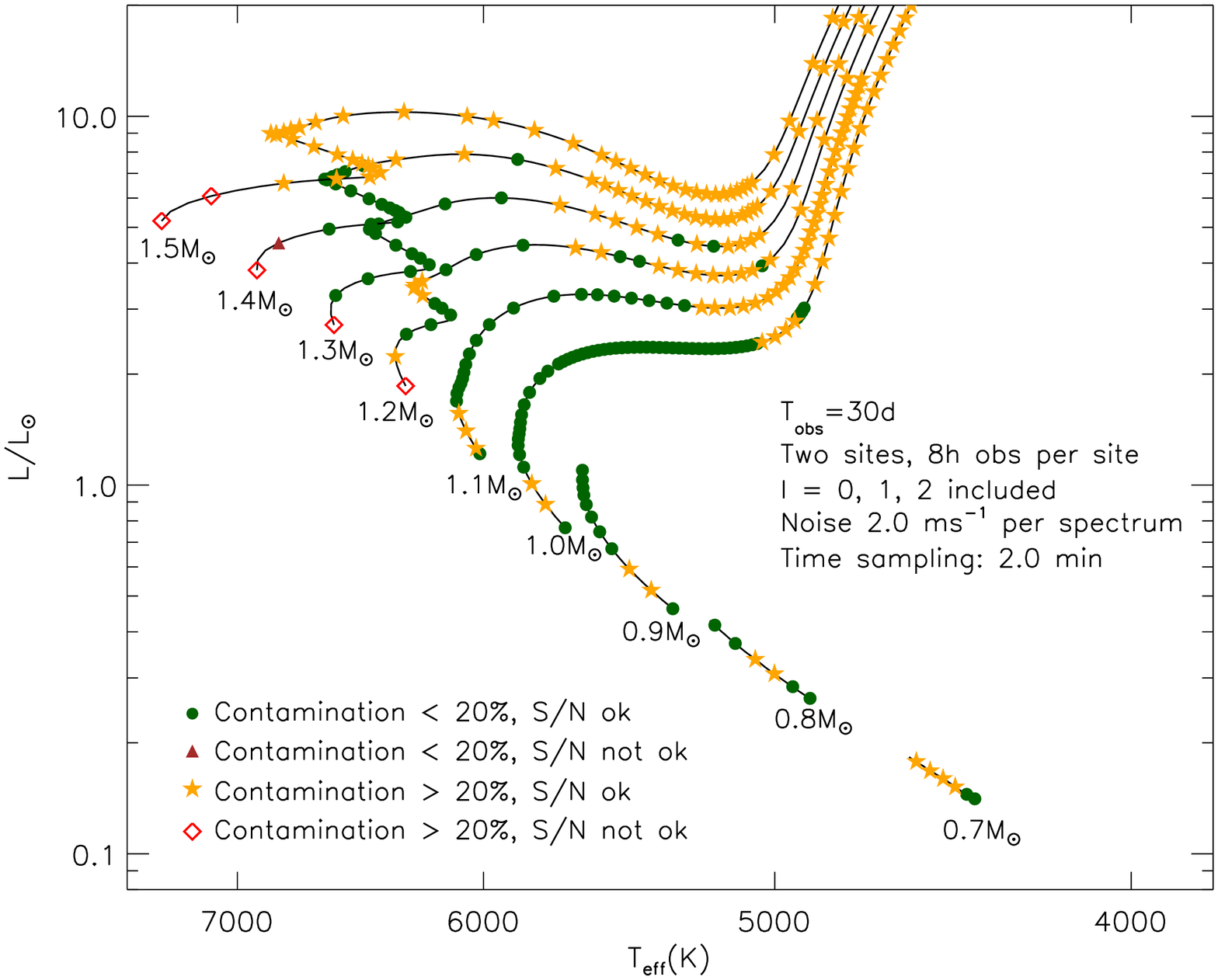} \\
\includegraphics[width=70mm]{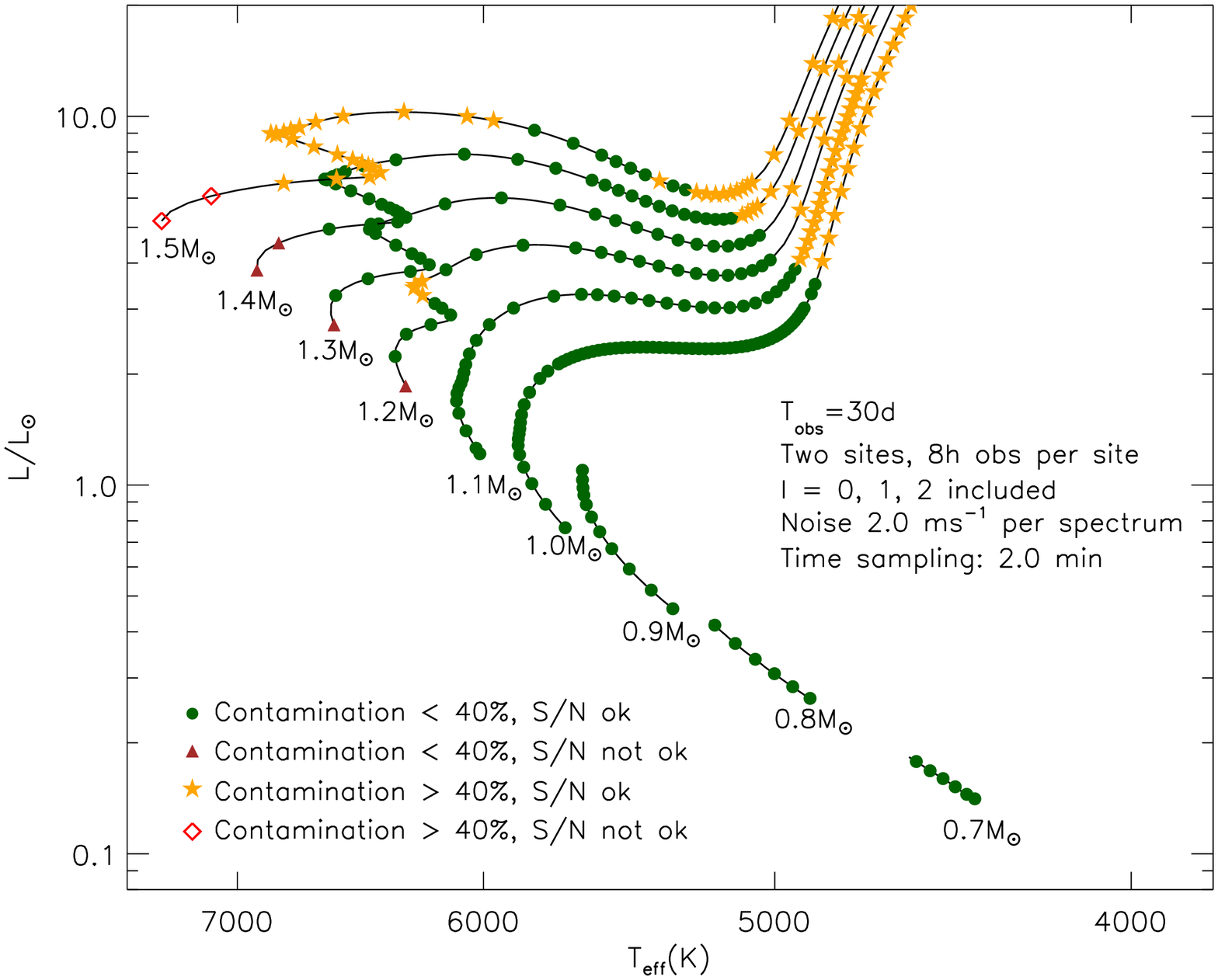} \\
\includegraphics[width=70mm]{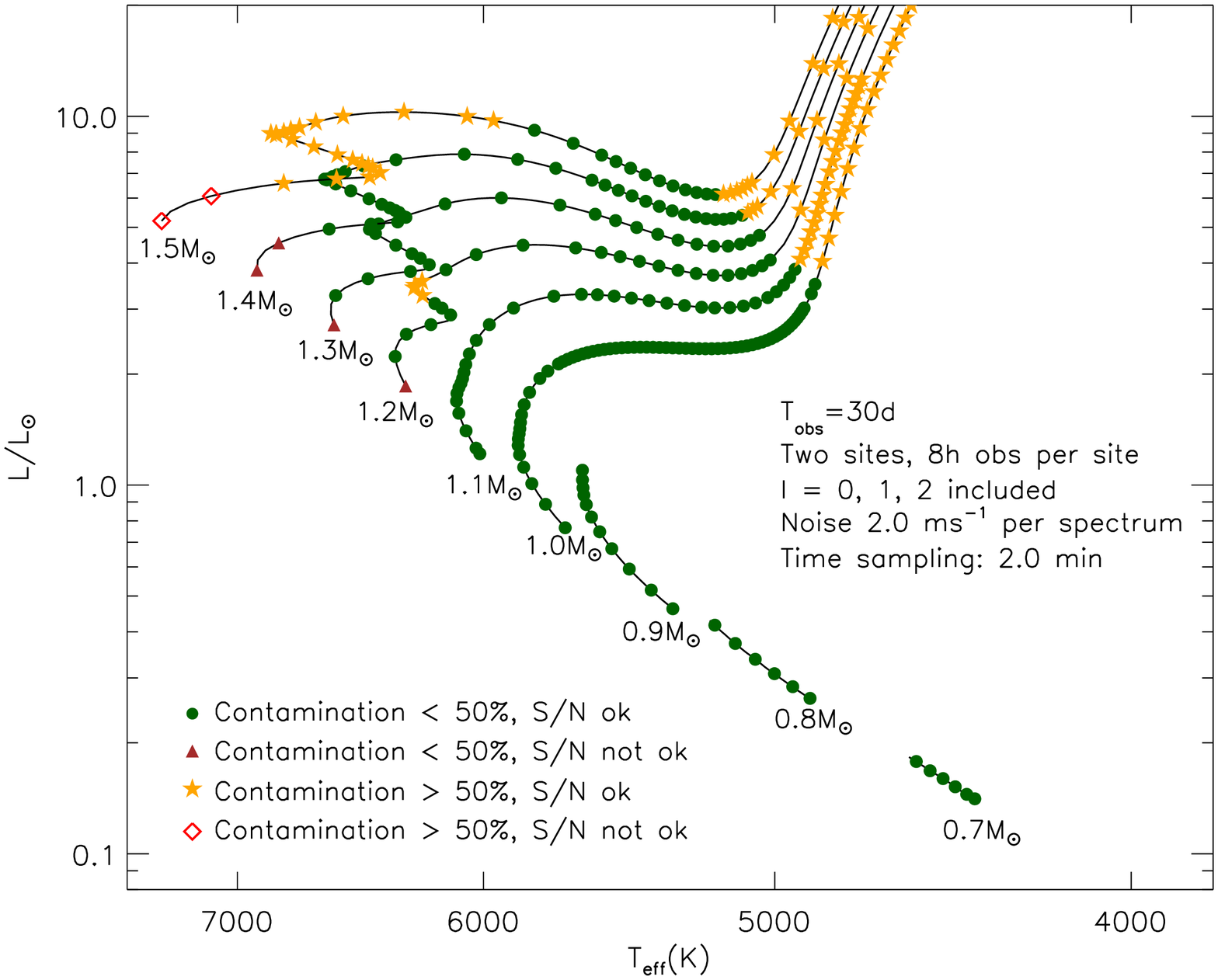} \\
\caption{Same as Fig.~\ref{fig.surfaceSS}, for two sites including 
$l=0-2$ and acceptable 
contamination levels of 10, 20, 40 and 50 per cent.} 
\label{fig.DS2}
\end{figure}

\begin{figure}
\includegraphics[width=70mm]{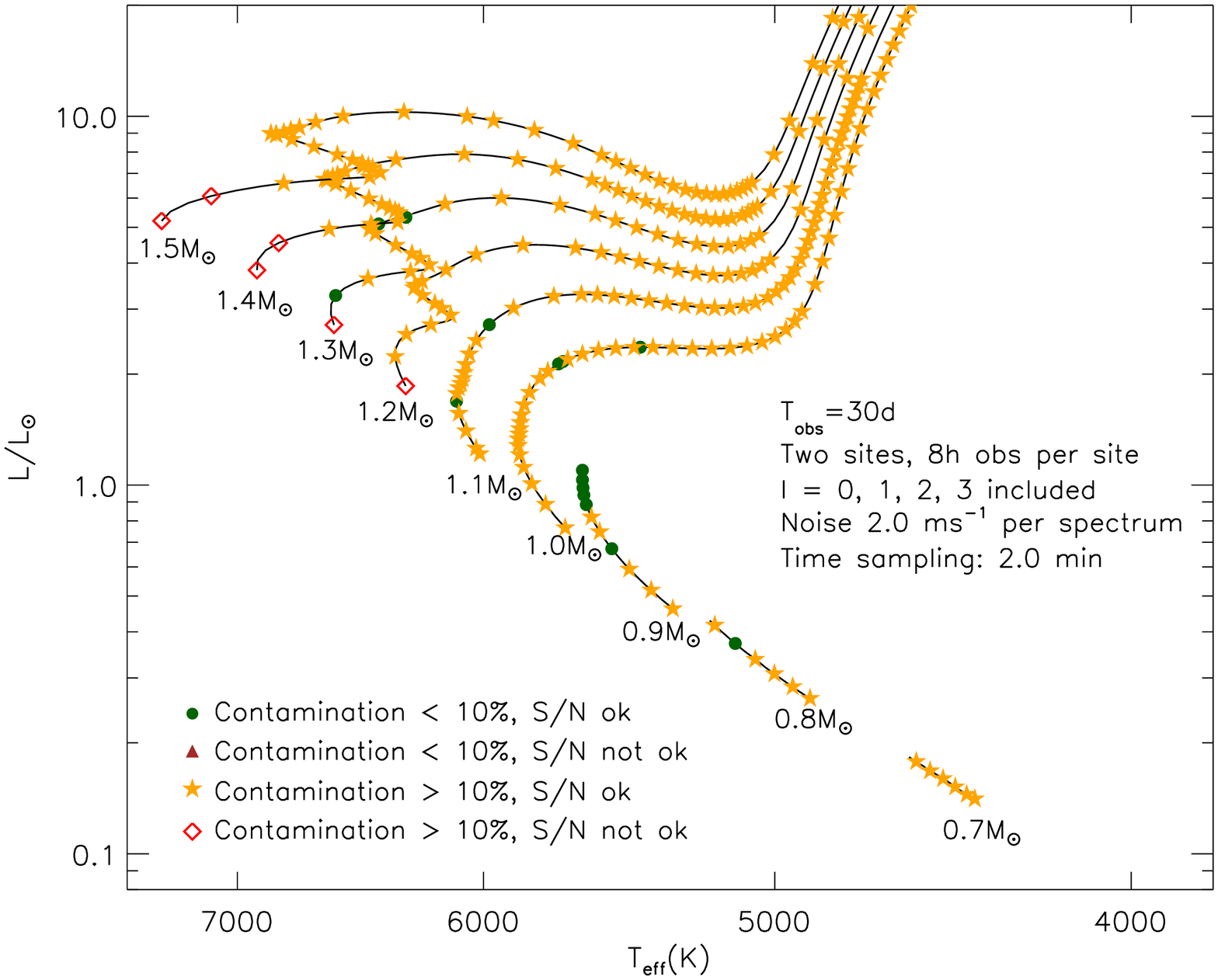} \\
\includegraphics[width=70mm]{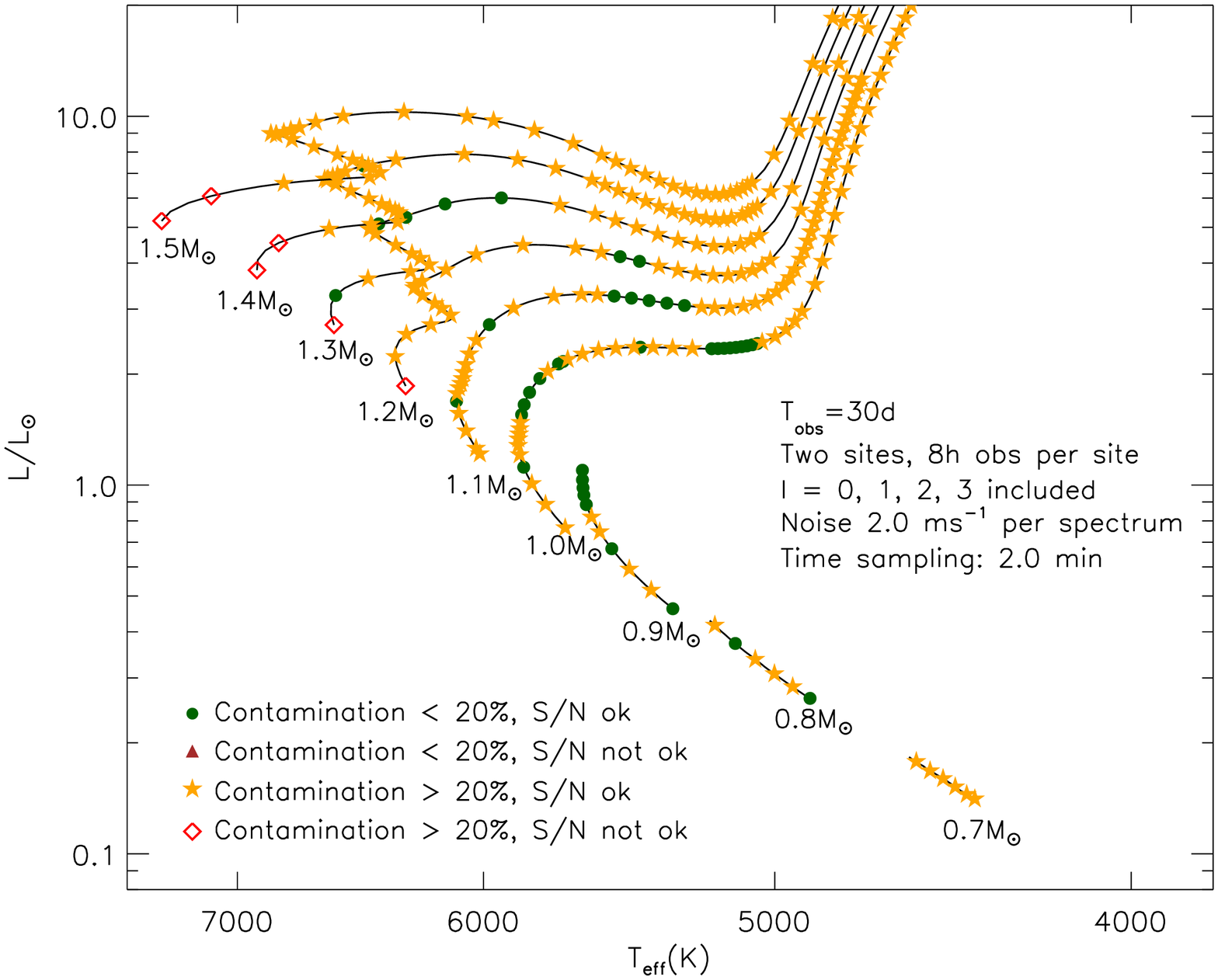} \\
\includegraphics[width=70mm]{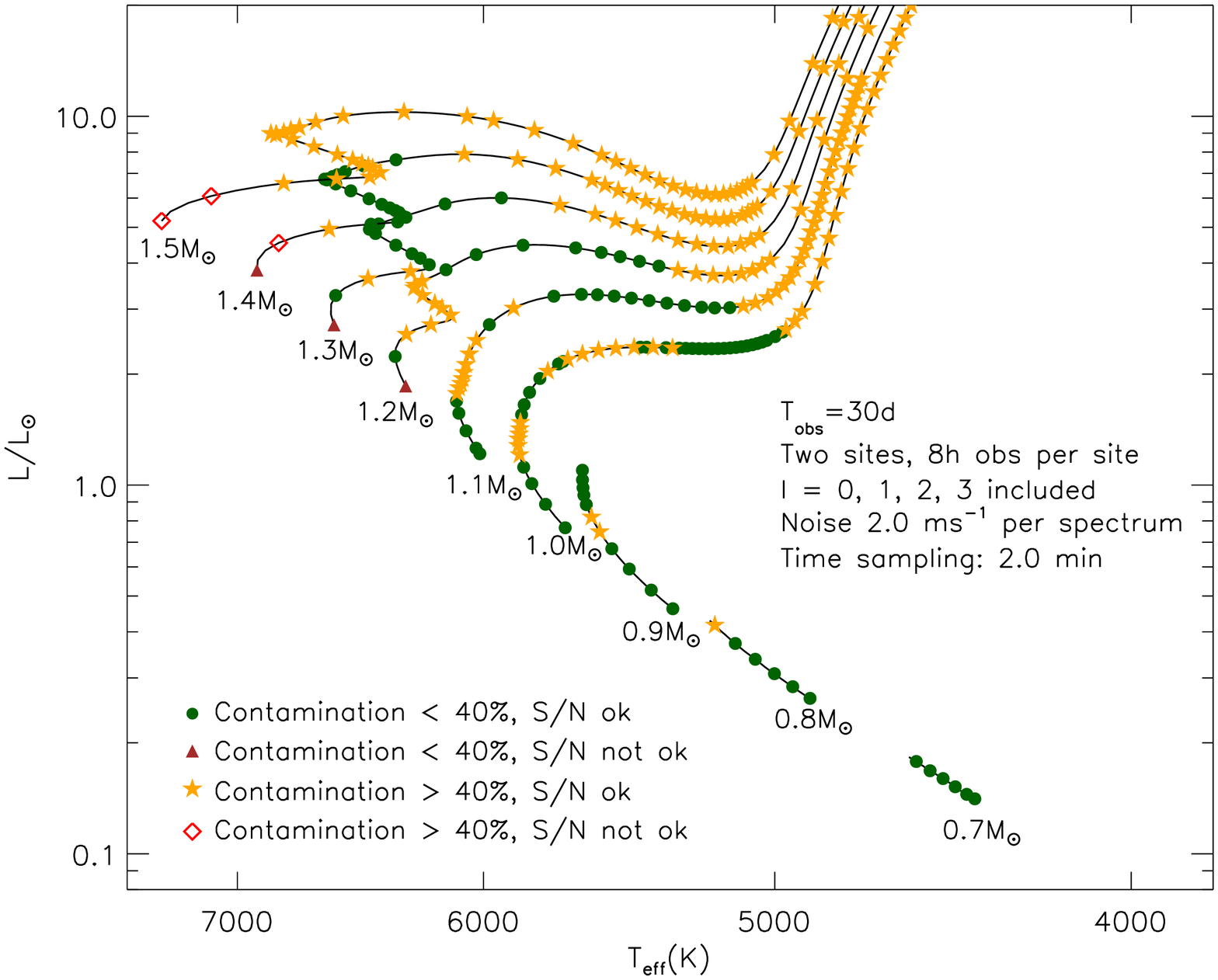} \\
\includegraphics[width=70mm]{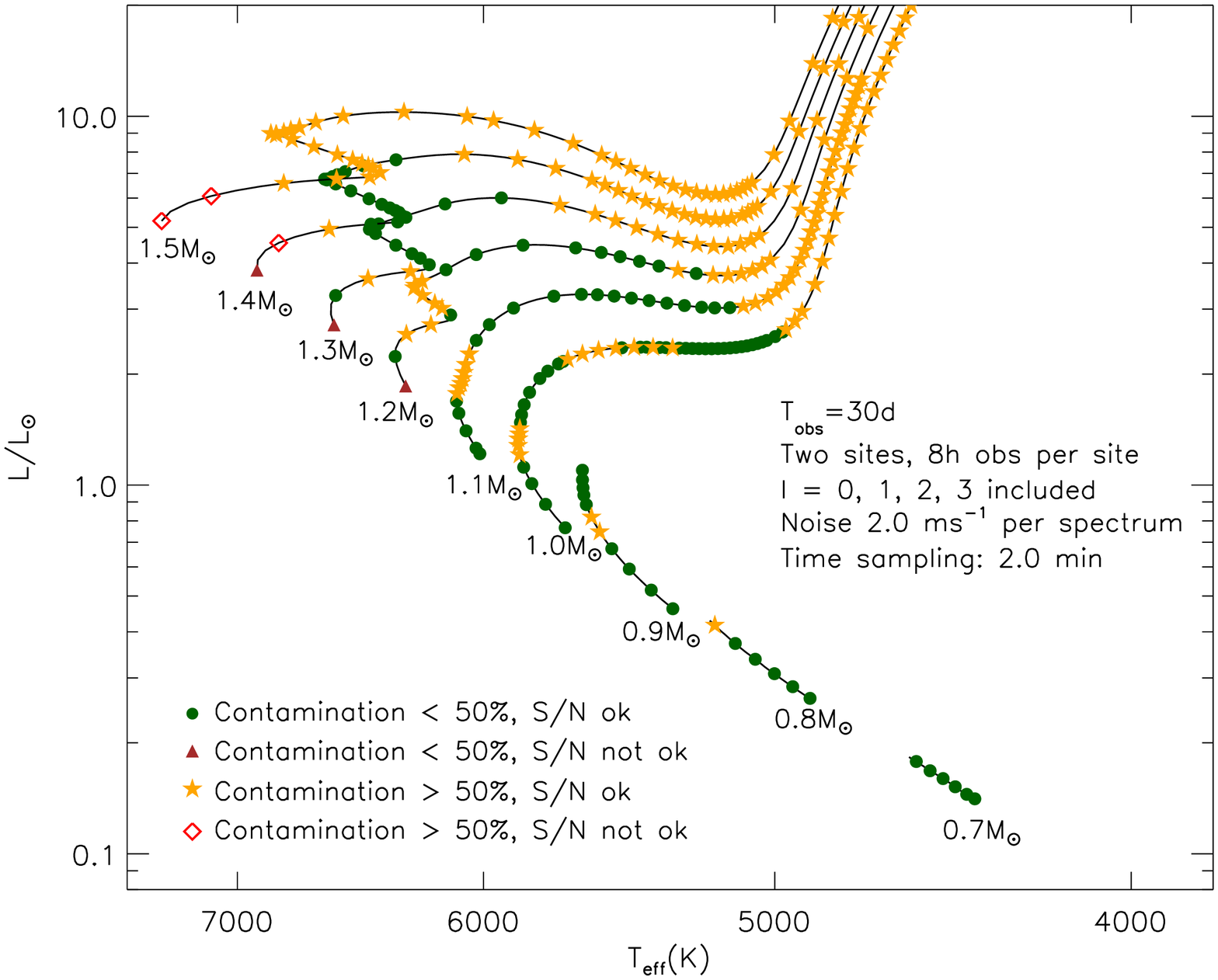} \\
\caption{Same as Fig.~\ref{fig.surfaceSS}, for two sites including 
$l=0-3$ and acceptable 
contamination levels of 10, 20, 40 and 50 per cent.} 
\label{fig.DS3}
\end{figure}

\begin{figure}
\includegraphics[width=70mm]{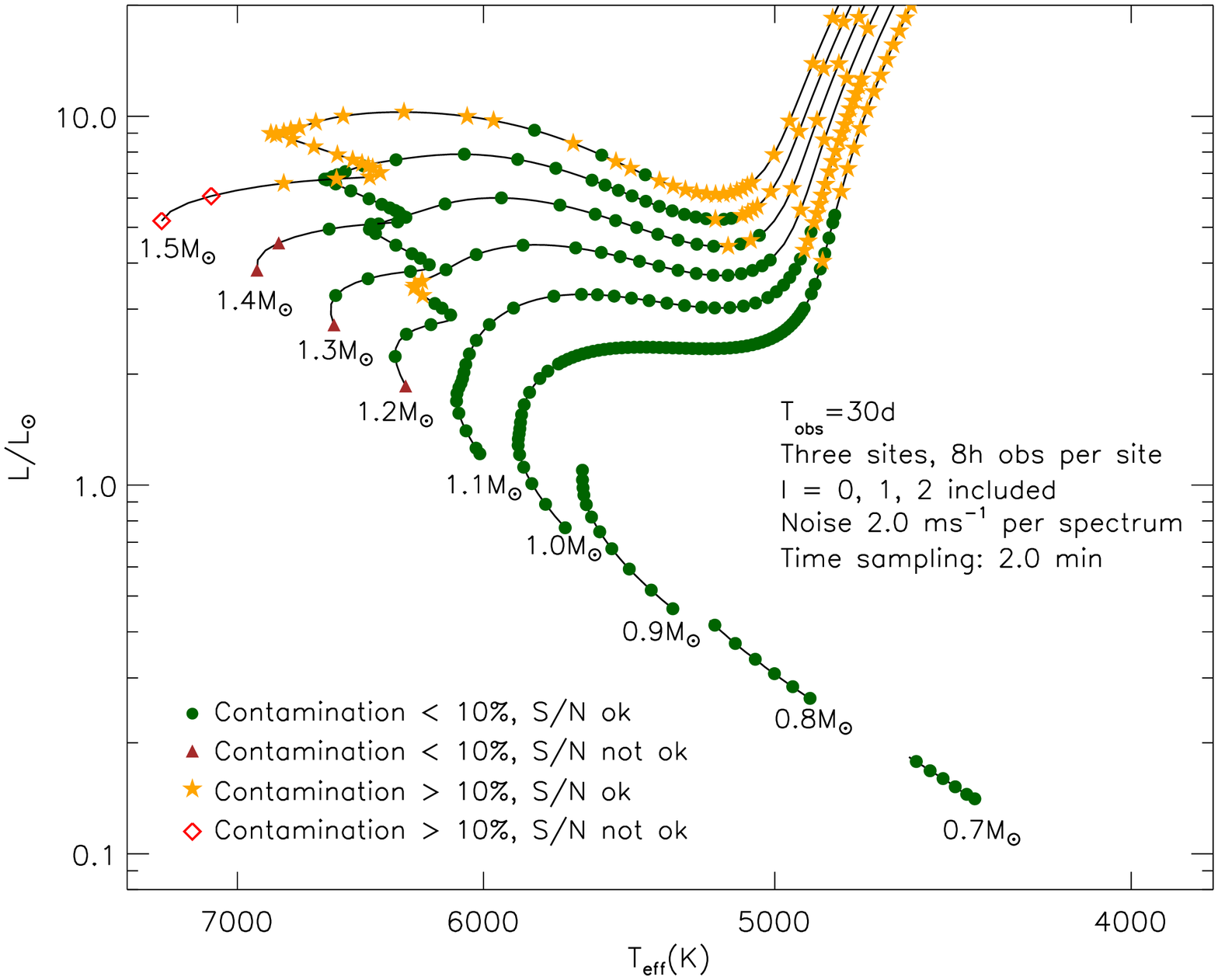} \\
\includegraphics[width=70mm]{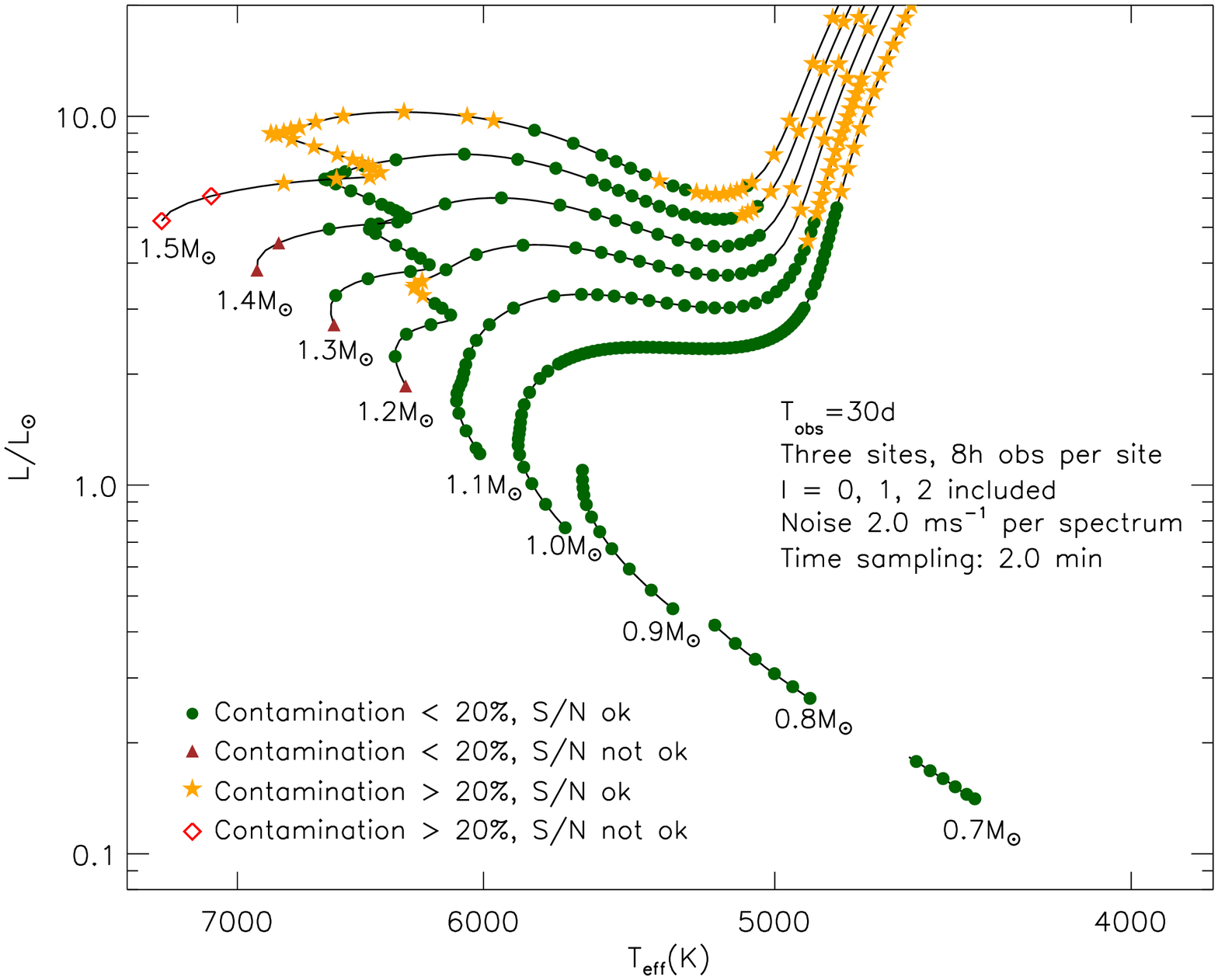} \\
\includegraphics[width=70mm]{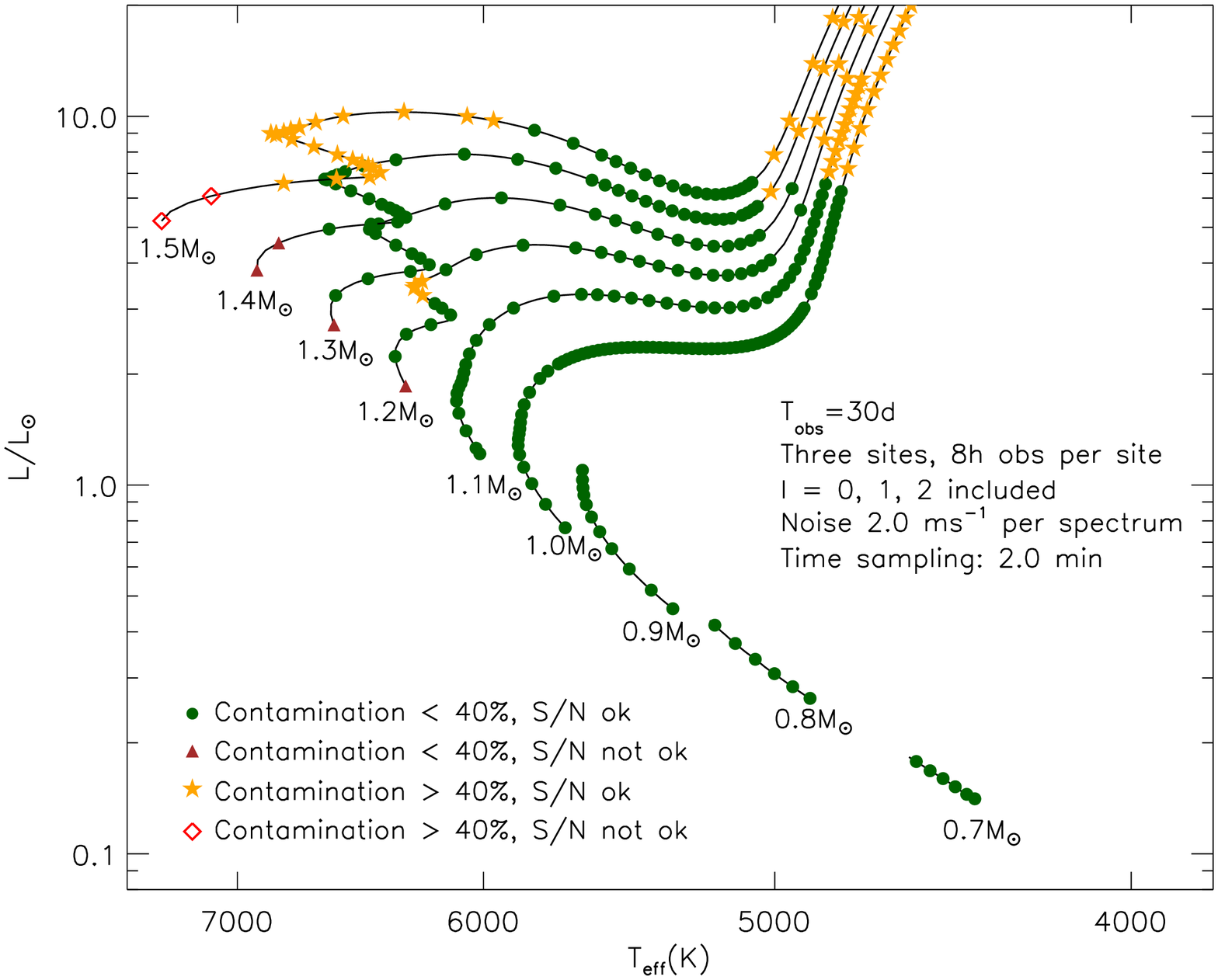} \\
\includegraphics[width=70mm]{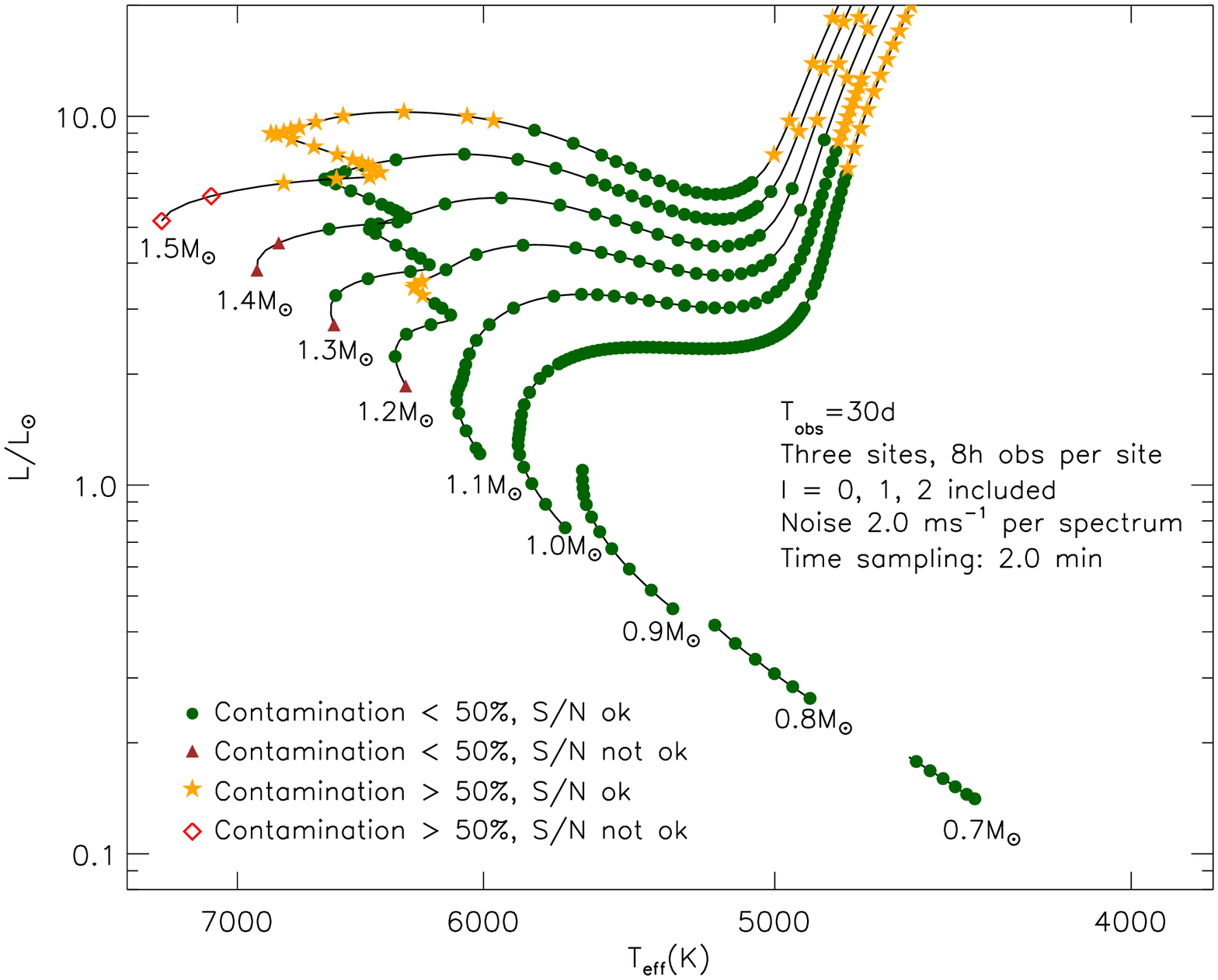} \\
\caption{Same as Fig.~\ref{fig.surfaceSS}, for three sites including 
$l=0-2$ and acceptable 
contamination levels of 10, 20, 40 and 50 per cent.} 
\label{fig.TS2}
\end{figure}

\begin{figure}
\includegraphics[width=70mm]{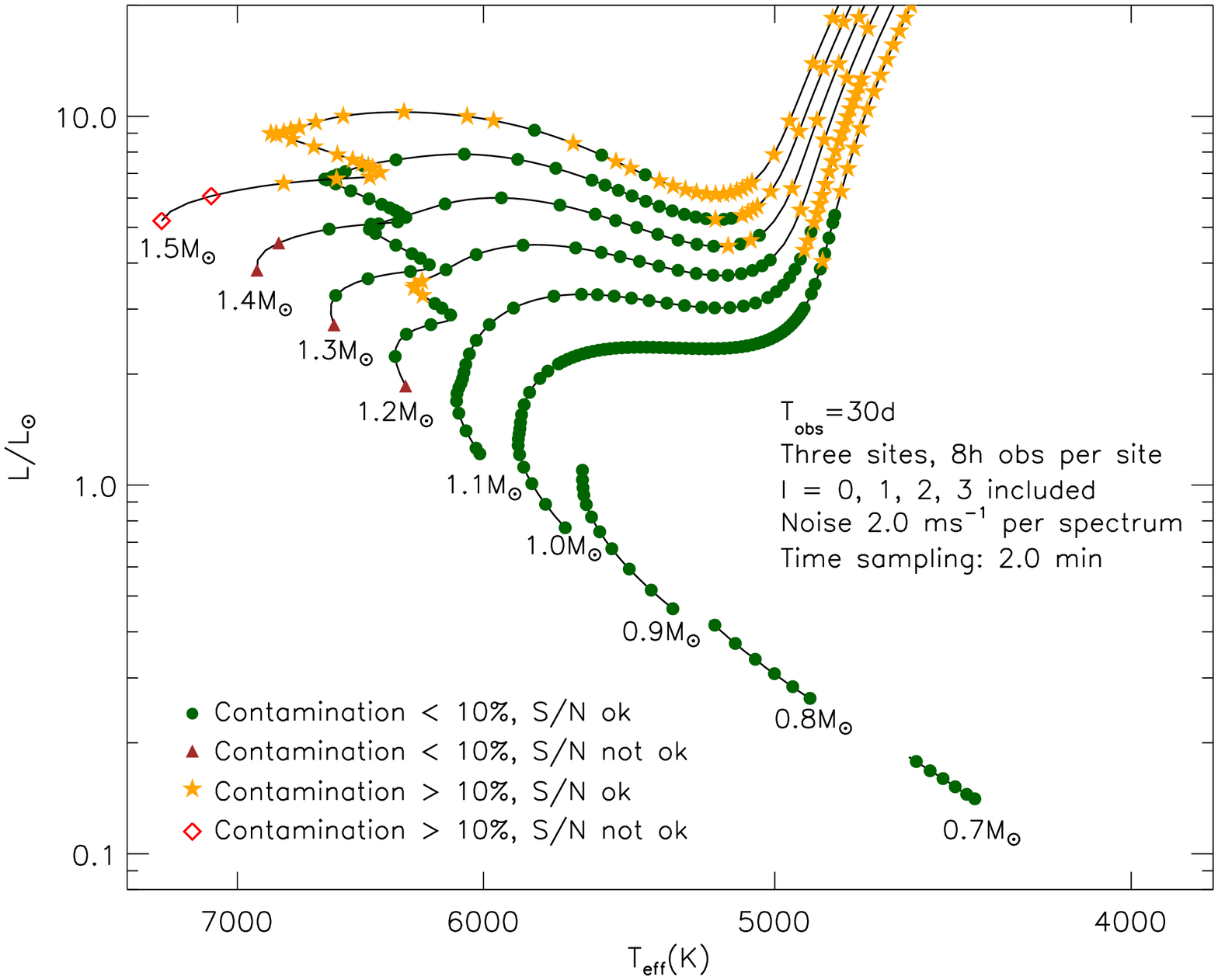} \\
\includegraphics[width=70mm]{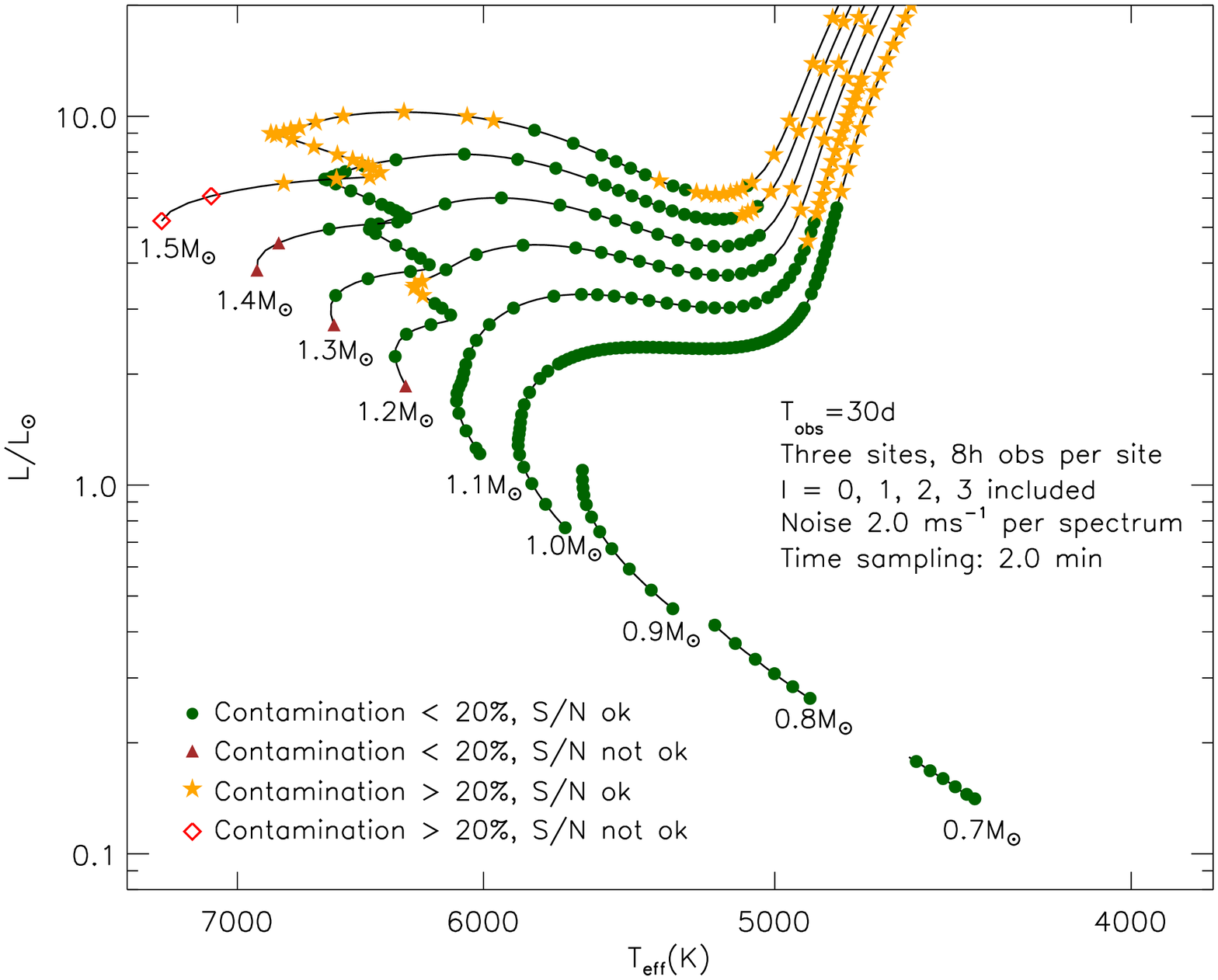} \\
\includegraphics[width=70mm]{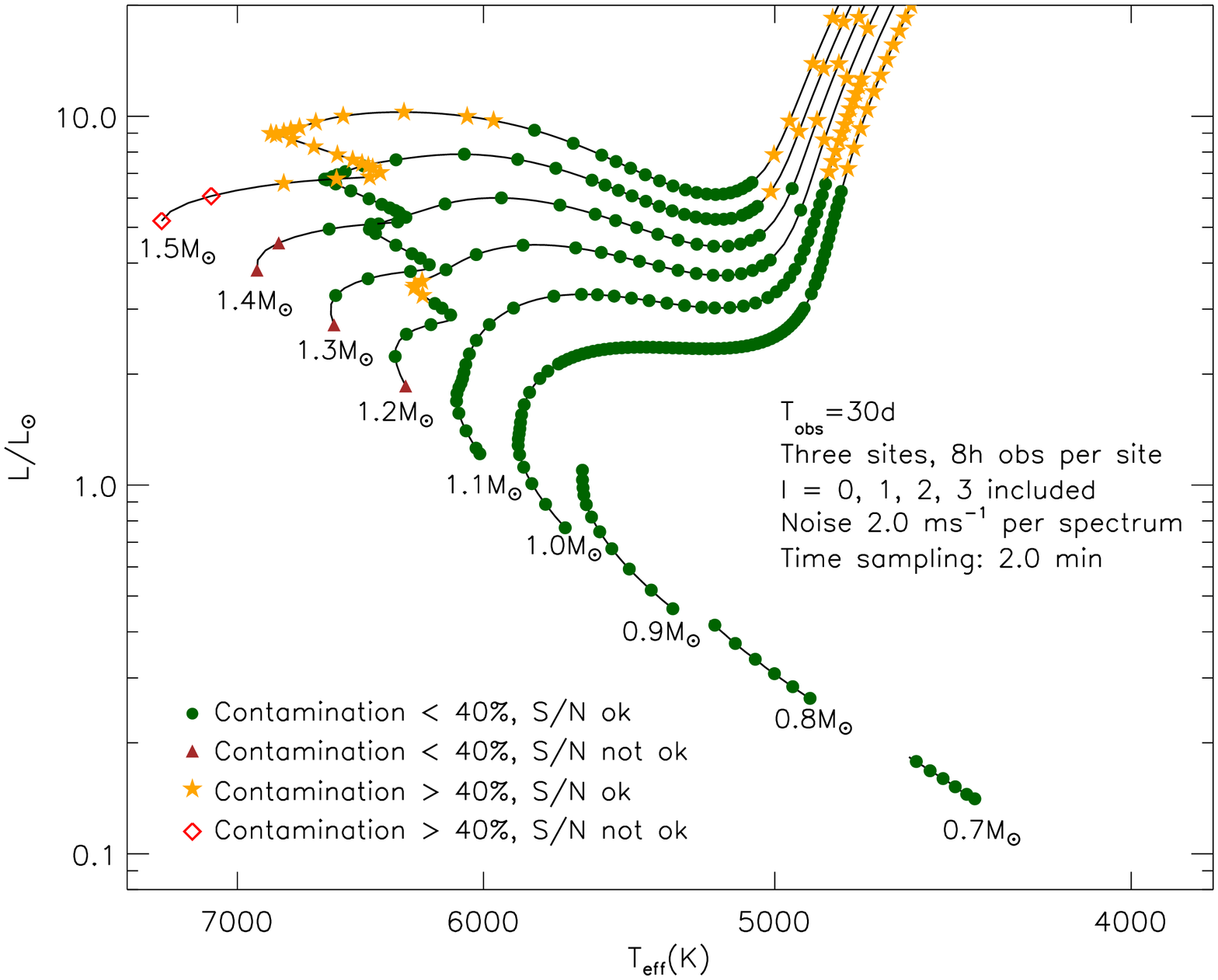} \\
\includegraphics[width=70mm]{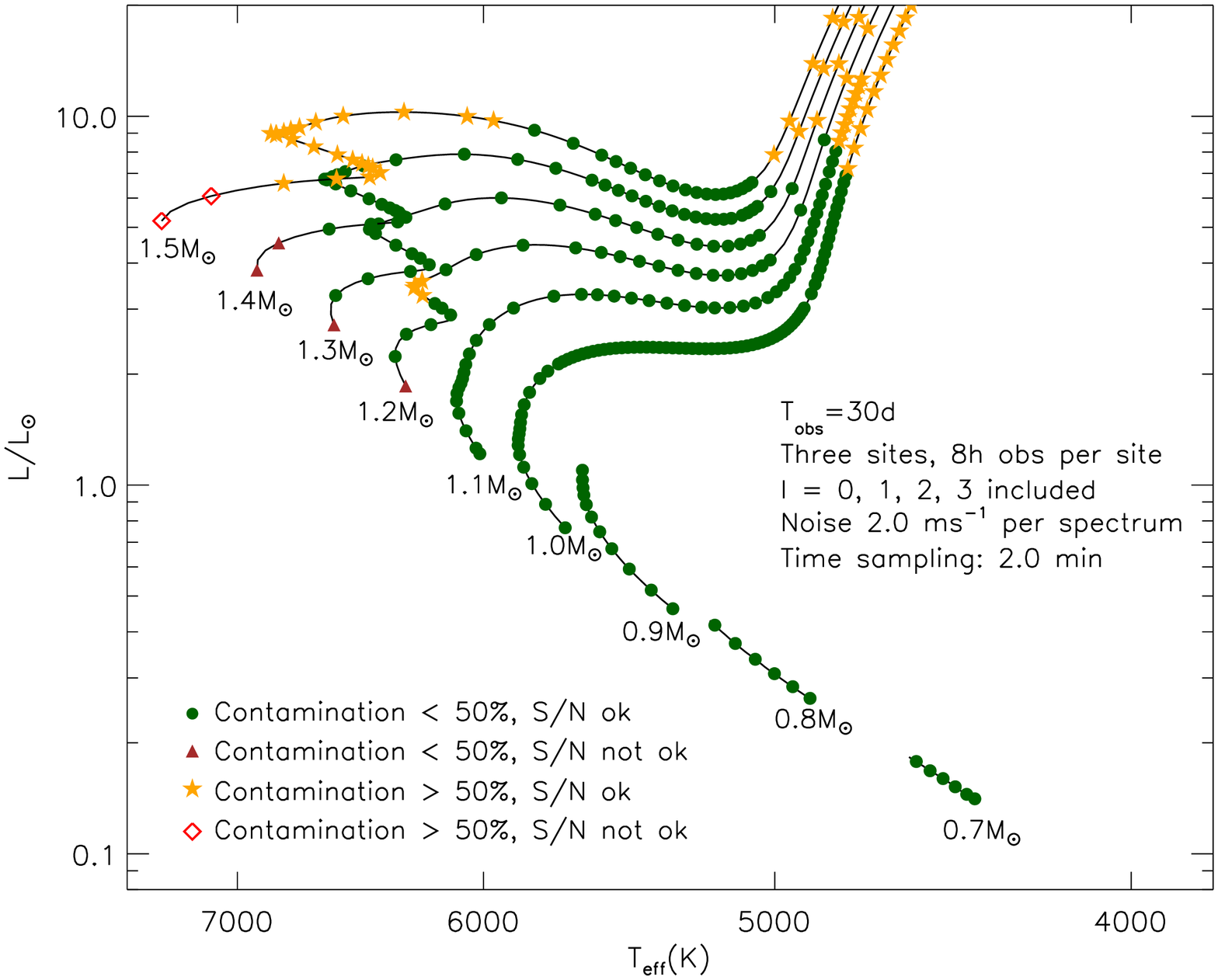} \\
\caption{Same as Fig.~\ref{fig.surfaceSS}, for three sites including 
$l=0-3$ and acceptable 
contamination levels of 10, 20, 40 and 50 per cent.} 
\label{fig.TS3}
\end{figure}

Figs.~\ref{fig.codedSS}--\ref{fig.coded3DS} contain the results
of the extended analysis, which includes realistic amplitudes and linewidths.
Instead of being plotted in a C-D diagram, the evolutionary tracks are plotted
in an HR diagram, showing in which 
regions of the HR diagram solar-like oscillations can be detected and
analysed and in which regions of the diagrams the analysis will be 
limited by too low signal-to-noise or by mode overlap. 
The latter can be caused by either a non-optimal combination of frequency 
splittings and spectral window function or by large mode linewidths. 
The plots shown here are for a moderate accepted overlap between 
frequency peaks of up to 30\% in amplitude. Similar diagrams can 
be made for the 4 other thresholds we have used; 10\%, 20\%, 40\% and 50\%.
These plots are shown in Figs.~\ref{fig.SS2}--\ref{fig.TS3}.

These plots show that significant, but not all, parts of the 
HR diagram are excluded for asteroseismic analysis when observing with a 
single SONG telescope, as well as when observing with two. 
With three SONG-nodes, seismic analysis is possible across most parts
of the HR diagram where 
solar-like oscillations are expected to occur. For the more massive and hot 
stars, however, a large mode linewidth combined with relatively low oscillation 
amplitudes will limit the analysis, even for a fully deployed SONG network. 

Fig.~\ref{fig.coded3DS} is much like 
Figs.~\ref{fig.codedSS}--\ref{fig.codedTS} but includes the $l=3$
modes, which are lower amplitude than the $l=0,1,2$ modes.
Fig.~\ref{fig.coded3DS} shows the case of two sites; the corresponding 
figure for three sites is identical to Fig.~\ref{fig.codedTS}.
We do not show the figure for a single site and including $l=3$; in this 
case, most stellar models are excluded due to mode overlap.
This is even the case for two sites (Fig.~\ref{fig.coded3DS}), where the 
majority of the models are excluded.
Adding a third node expands the possibilities considerably,
however (see Fig.~\ref{fig.codedTS}).

\section{Discussion}

The analysis presented above can not only be used for evaluating target stars 
for observing solar-like oscillations with a network like SONG, but it 
can also be used for optimizing observations with such a network. 

\begin{figure}
\includegraphics[width=98mm]{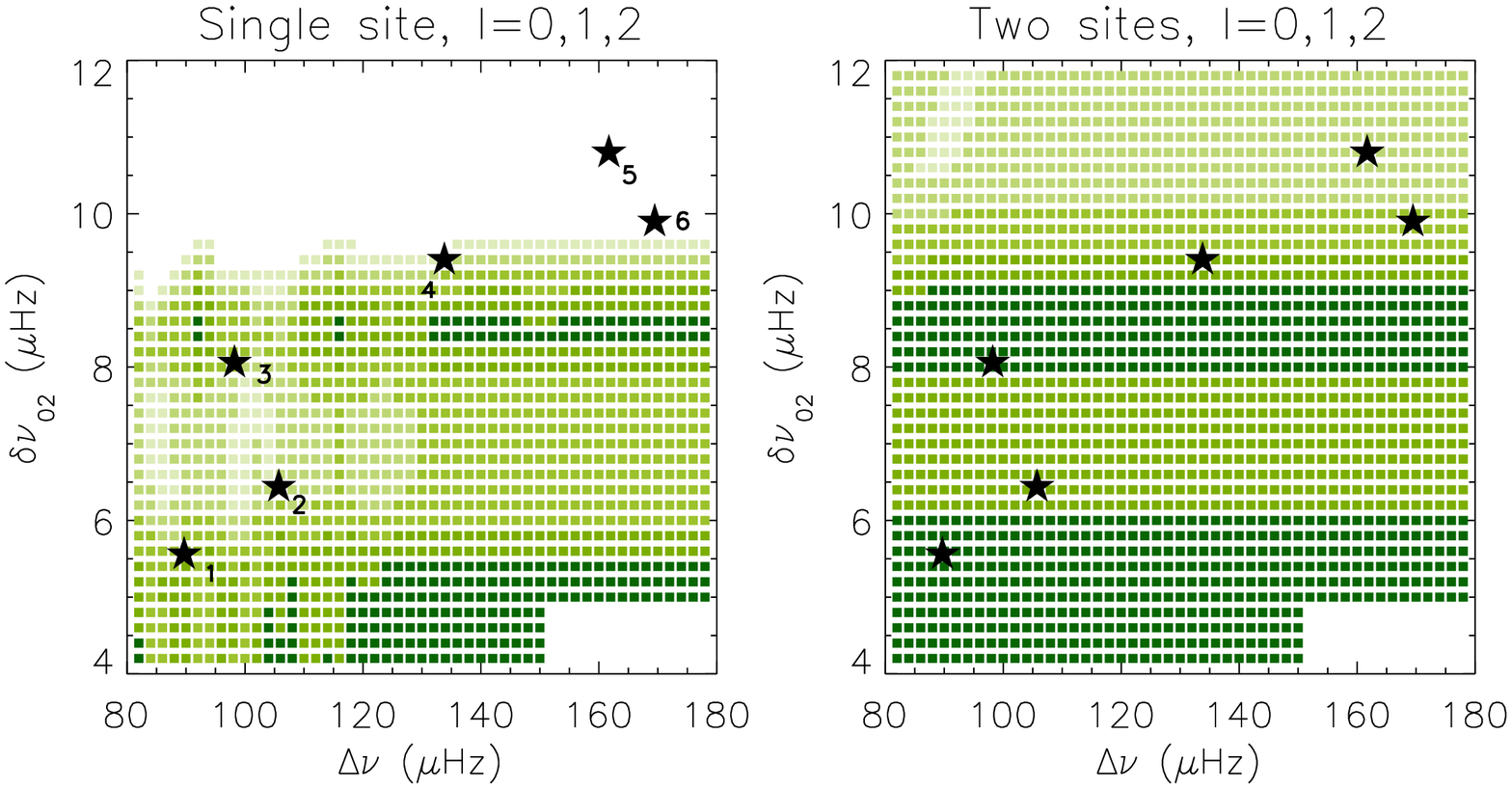} \\
\includegraphics[width=98mm]{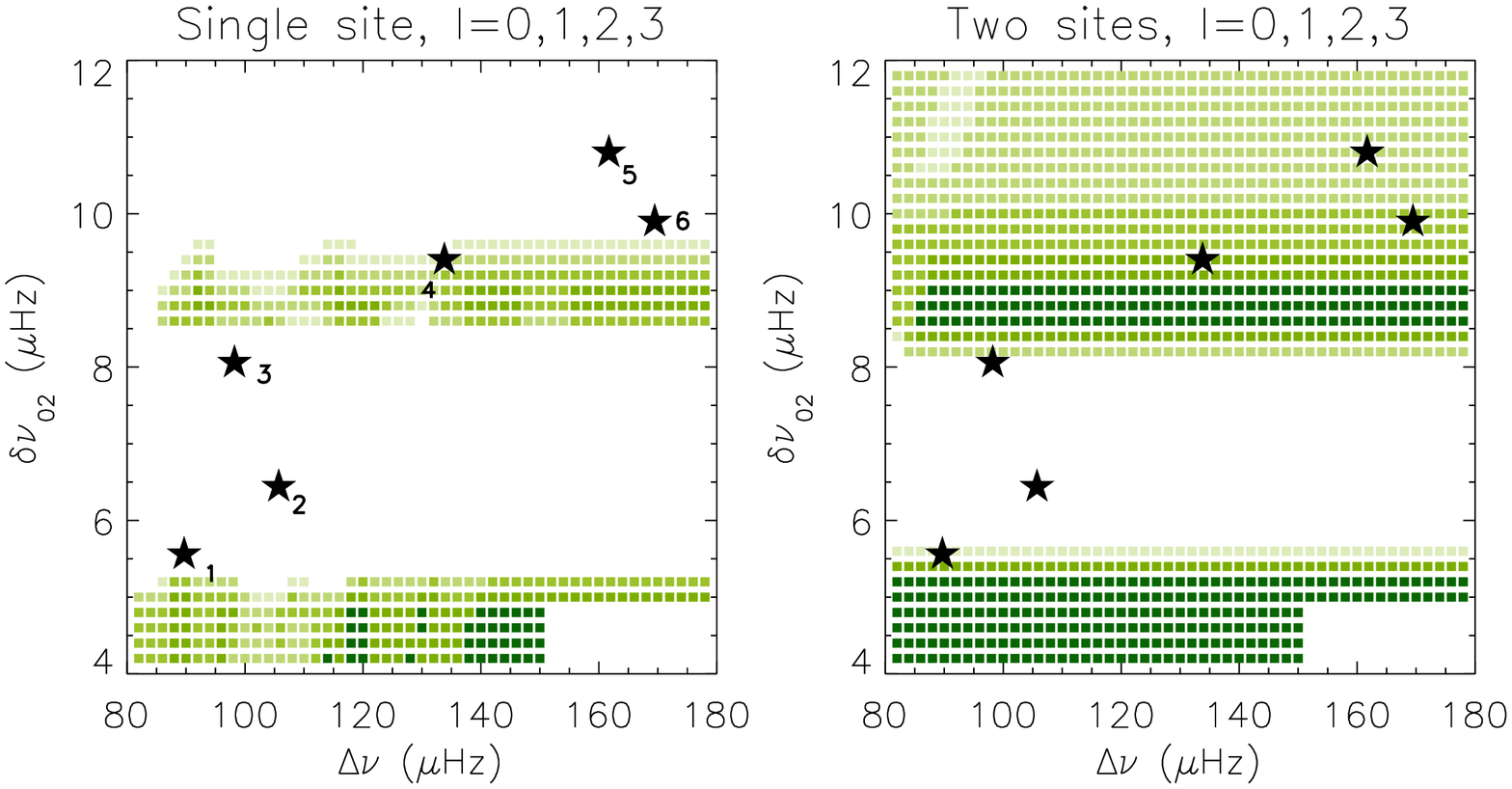}
\caption{
Sample diagrams demonstrating the analysis of individual
targets for network of telescopes like SONG. 
The panels show the contamination level for 
one and two SONG nodes, without (top) and with (bottom) $l=3$ included
for 6 well-studied stars:
1: $\mu$\,Ara \citep{b1}; 2: $\alpha$\,Cen A \citep{b3}; 
3: HD52265 \citep{b4}; 4: 18\,Sco \citep{b5} (and the Sun); 
5: $\alpha$\,Cen B \citep{kb2}; 
6: $\tau$\,Ceti \citep{t2}. Colour-coding as in Figs.~\ref{fig.surfaceSS}--\ref{fig.surfaceDS-l3}. 
}
\label{fig.targets}
\end{figure}

In Fig.~\ref{fig.targets} we show the level of contamination due to mode
overlap for 6 well-observed
stars for which the frequency separations of solar-like oscillations have 
been measured. It is evident that a star similar to $\alpha$\,Cen B, which has a 
small frequency separation $\delta\nu_{02}$ near 11\,$\mu$Hz, is not a good 
target for single site observations aiming at determining the small frequency 
separation. 

From the upper left-hand panel of Fig.~\ref{fig.targets}, we can identify the types
of stars for which a single node will be sufficient: those lying below and
to the right of 18\,Sco, as the dark-green areas indicate regions for which
the acoustic spectra of stars will not be 
significantly affected by overlap between different mode peaks. This complements
the information in Fig.~\ref{fig.codedSS}, which indicates that the amplitudes and
mode linewidths of such stars are suitable for study with a single SONG-node
for the noise level we have assumed.

The upper right-panel of Fig.~\ref{fig.targets} represents a network with
two nodes. In comparison to a single node, the number of stars for which 
contamination is a problem is significantly reduced. 
Consequently, we can envision studying multiple stars at a time with a
fully deployed SONG network, with perhaps 4 telescopes 
distributed on the northern and southern hemispheres, allocating
two nodes per target and doubling the rate at which SONG measures
the solar-like oscillations in stars.

\begin{figure}
\includegraphics[width=98mm]{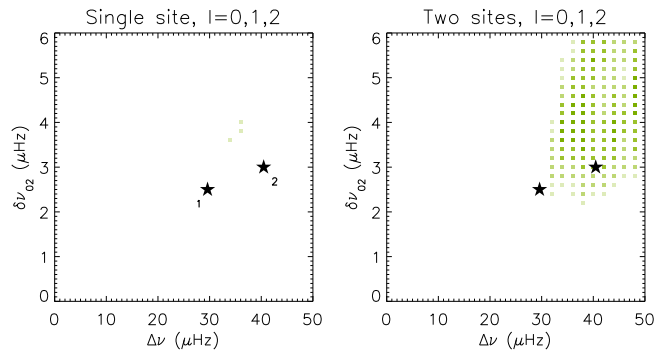}
\caption{
Same as Fig.~\ref{fig.targets}, except that two subgiants 
for which stellar oscillations have been observed are included:
1: $\beta$\,Aql \citep{cos1}; 2: $\eta$\,Boo \citep{kb3}. 
}
\label{fig.subGi}
\end{figure}

Sub-giant stars have comparatively high oscillation amplitudes and are 
therefore potentially good targets for observations with either 
one or two SONG nodes. Fig.~\ref{fig.subGi} depicts a subsection of the 
C-D diagram, where sub-giants are found. It can be seen that for some 
sub-giants, as for example $\eta$ Boo, the combination of the small and 
the large frequency separations will allow seismic analysis without 
significant contamination from the spectral window function, when observing 
with two SONG nodes. When observing from a single SONG node, the analysis 
will be strongly affected by effects of the spectral window function. 

The results presented in Figs.~\ref{fig.codedSS}--\ref{fig.coded3DS}
might erroneously suggest that we shall not be able to observe
oscillations in red giants with SONG.
In fact, the present analysis is based on 
the asymptotic relation and other assumptions, as described in Sect.~2,
which are not entirely relevant for red giants.
In particular, the lower frequencies and consequently smaller frequency
separations for these stars mean that observations longer than the 30\,d 
assumed here will be needed to avoid confusion.
A separate analysis will be required to evaluate the potential of SONG for
the study of red giants.

\section{Conclusions}

The methods presented here demonstrate an approach for evaluating targets
for measuring solar-like oscillations spectroscopically using either a single
telescope or a network of telescopes. While the SONG network is the inspiration
for this analysis, it is generalizable to any single or multi-site campaign
with a similar goal in mind.

We have shown that under the assumptions we have made, for certain targets, 
one or two SONG telescopes will be sufficient to observe solar-like 
oscillations without the subsequent seismic analysis being hampered by 
overlapping mode peaks in the amplitude spectrum. 
However, it is important to remember that the specific science goals of
an individual project define the acceptable level of contamination
for an amplitude spectrum or what $S/N$-level is required -- or is even
possible. It is not the goal of this paper to recommend or rule out
potential solar-like targets for SONG.
This analysis is intended to assess the benefit of additional nodes
in the SONG network and to develop a tool for future
evaluations of specific targets for SONG, 
taking information like target brightness, rotational velocity and position 
in the HR diagram and contributing to the target selection process and scheduling for SONG.

Based on our analysis, we conclude that certain targets will require only one or
two SONG nodes to enable the detection of solar-like oscillations despite
contamination of the amplitude spectrum due to the window function. However,
the detection of solar-like $l=3$ modes requires full network coverage for most of
that part of the HR diagram where we expect stars to exhibit solar-like oscillations.
Given that detection of $l=3$ 
modes is an important science driver for SONG -- 
and one of the factors that will push the 
analysis of solar-like stars beyond what can be reached with the {\it Kepler} 
mission -- we conclude that, while the two SONG nodes currently under construction 
will represent an important advance for ground-based asteroseismology,
they are not sufficient to fulfill the ultimate science goals of SONG.

\section*{Acknowledgments}

Funding for the Stellar Astrophysics Centre is provided by The Danish
National Research Foundation (Grant DNRF106).
The research is supported by the ASTERISK 
project (ASTERoseismic Investigations with SONG and {\it Kepler}) funded by 
the European Research Council (Grant agreement no.: 267864).
We thank D. Stello for providing the model grid used in the analysis.

\label{lastpage}

\end{document}